\documentclass[preprint,12pt]{elsarticle}
\usepackage[dvipsnames]{xcolor}

\usepackage{amssymb}
\usepackage{amsmath}

\journal{Sustainable Energy and Fuels}

\begin{document}

\begin{frontmatter}



\title{A DFT study on 18-crown-6-like-N$_8$ structure as a  material for metal-ions storage: stability and performance}

\author{Irina I. Piyanzina $^{a,b*}$, Regina M. Burganova$^{b}$, Sadegh Kaviani$^{b}$, Oleg V. Nedopekin$^{b}$, Hayk Zakaryan$^{a}$  \\ 
irina.gumarova@ysu.am}

\affiliation{organization={Computational Materials Science Laboratory, Center of Semiconductor Devices and Nanotechnology, Yerevan State University, Republic of Armenia},    
addressline={1 Alex Manoogian St.}, 
           city={Yerevan},
            postcode={0025}, 
          country={Republic of Armenia}}

\affiliation{organization={Institute of Physics, Kazan Federal University},
            addressline={16~Kremlyovskaya~str.}, 
            city={Kazan},
            postcode={420008}, 
           country={Russia}}

\begin{abstract}
Developing electrode materials with exceptional electrical conductivity, robust chemical stability, rapid charge and discharge rates, and high storage capacity is essential for advancing high-performance metal-ion batteries. This study explores the two-dimensional, 18-crown-6-like N$_8$ structure (2D-N$_8$) as a promising electrode material for next-generation rechargeable post-lithium batteries.
We thoroughly investigated the pristine N$_8$ structures, focusing on their stability and performance metrics. Our analysis revealed remarkable structural stability across the board. Additionally, electronic calculations indicated a small band gap of 0.54\,eV for the N$_8$ monolayer, suggesting favorable electronic properties for battery applications.
When we evaluated a series of metal ions as adsorbates, we found that the pristine N$_8$ monolayer achieved an impressive storage capacity of 1675\,mAh/g for sodium (Na) and magnesium (Mg) ions, highlighting its potential for effective ion storage. 
Our findings suggest that the engineered 2D-N$_8$ structure offers a unique combination of stability and electrochemical performance that could significantly contribute to developing efficient and durable energy storage technologies.
\end{abstract}

\begin{graphicalabstract}
\includegraphics[width=\linewidth]{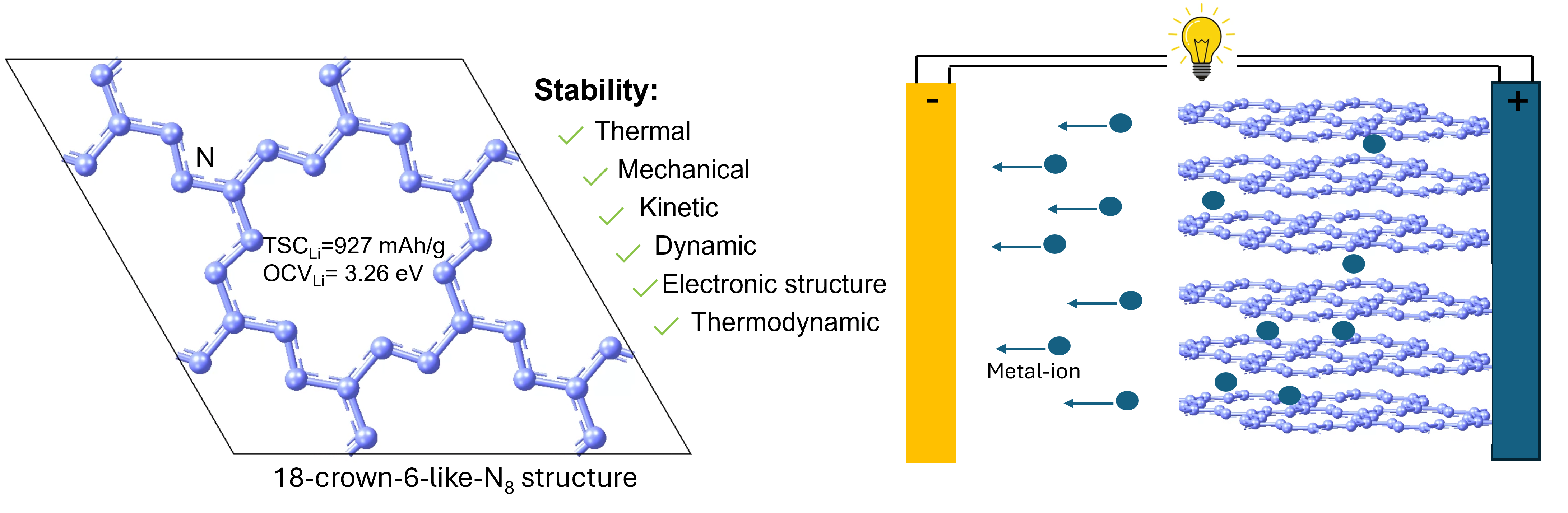}
\end{graphicalabstract}

\begin{highlights}
\item The N$_8$ monolayer exhibits good thermal and mechanical stability.
\item The N$_8$ monolayer is a small-band gap semiconductor.
\item Calcium (Ca) has the lowest adsorption energy at the N$_8$ monolayer.
\item The N$_8$ shows a storage capacity of 1675 mAh/g for sodium (Na) and magnesium (Mg) ions.
\item The lowest diffusion barrier is for potassium (K) ions.
\item The calculated OCVs (2.26-3.36\,eV) are suitable for cathode applications.

\end{highlights}

\begin{keyword}
2D material \sep high-energy material \sep crown-like structure \sep polynitrogen \sep metal adsorption  \sep DFT 
\end{keyword}

\end{frontmatter}


\section{Introduction}
\label{sec:intro}
Developing renewable and environmentally friendly energy sources is essential, as fossil fuel resources are finite and harmful to the ecosystem\,\cite{1,2,3}. Rechargeable batteries represent a promising choice in the current energy storage landscape, as they must meet high-energy-density, safety, and affordability requirements. Lithium-ion batteries (LIBs) were first introduced to the market in 1991 by Sony Corp., quickly gaining popularity due to their impressive features, such as extended cycle life and high energy density\,\cite{4,5}. The scarcity of lithium reserves restricts the application of lithium-ion batteries and drives their costs. Furthermore, the theoretical specific capacity of graphite, frequently used as the anode material in LIBs, is 372\,mAh/g, below the demand for most practical applications\,\cite{6}. Thus, the importance of post-lithium ion batteries, including sodium-ion batteries (SIBs)\,\cite{7}, potassium ion batteries (KIBs)\,\cite{8}, magnesium ion batteries (MgIBs)\,\cite{9}, and calcium ion batteries (CaIBs)\,\cite{10} should not be overlooked. Sodium and potassium are located directly below lithium in the periodic table, which means these three elements have similar chemical properties. Their abundance in Earth's resources and lower extraction cost compared to lithium make them promising candidates for use in next-generation battery systems\,\cite{11}. Magnesium and calcium batteries tend to have a more stable electrochemical profile, which can lead to improved safety\,\cite{12,13}. Because they usually operate at lower standard electrode potentials, the risk of dendrite growth may be lower. The electrochemical performance of post-lithium ion batteries at the electrodes is somewhat limited due to the larger atomic radii of post-lithium metals, resulting in a relatively lower storage capacity, poorer ionic diffusion, and larger volume changes in the anode\,\cite{14}. Consequently, designing new electrode materials with high energy density, storage capacity, and ion mobility is crucial while exhibiting negligible expansion and contraction for post-lithium-ion batteries.

Two-dimensional (2D) materials exhibit significant advantages as potential cathode candidates compared to conventional materials\,\cite{15}. This is because of their extensive specific surface area, multiple adsorption sites, and outstanding mechanical strength. Currently, a variety of abundant 2D electrodes, including borophene\,\cite{16}, silicene\,\cite{17}, phosphorene\,\cite{18}, transition metal dichalcogenides (TMDs)\,\cite{19}, and MXenes\,\cite{20} have been studied for their application in LIBs.  However, many of these materials have inherently certain limitations. For example, borophene and silicene are chemically unstable in air and oxygen, leading to degradation over time\,\cite{21,22}. This instability can negatively affect the overall lifespan and performance of the battery. Although black phosphorene exhibits excellent rate performance and a high specific capacity, significant volume variation during the charge and discharge processes hinders its practical use\,\cite{23}. Incorporating transition metals tends to increase the molar mass of electrode materials, potentially reducing the specific capacity. WSe$_2$ and MoS$_2$ as transition metal dichalcogenides suffer from low electronic conductivity, thus deteriorating their low charge mobility. Finally, the adsorption capacity and diffusion barrier of multivalent ions are significantly influenced by the terminal groups on MXenes, which affect the theoretical capacity and the charge/discharge rates of batteries\,\cite{24}. Therefore, there is a strong demand for 2D electrode materials that demonstrate exceptional overall performance. Developing such materials for metal-ion batteries continues to be a significant challenge. 

Polynitrogen-containing materials, characterized by high nitrogen content and unique structural properties, have attracted increasing attention in energy storage and conversion due to their high energy density and ease of electron transfer\,\cite{25}. However, one of the notable drawbacks of polynitrogen-containing materials is their instability. A proven approach to improving the stability of polynitrogen compounds is to create a nitrogen network that incorporates both N-N and N=N bonds and to develop nitrogen-rich alloys with other elements\,\cite{26}. Strengthening the interactions between metal cations and anionic polynitrogen structures can lead to the development of novel high-energy density materials. Recent advances in synthetic methodologies have facilitated the design and optimization of polynitrogen compounds, revealing their intriguing physicochemical properties, such as high stability, energetic output, and tunable reactivity\,\cite{27,28,29,30}.

In 2017, by Steele and Oleynik\,\cite{47} the new P6/mcc KN$_8$ compound was predicted at 70\,GPa by evolutionary algorithms and the methodology of functional density theory. 
More recently, Wang et al.\,\cite{uspexn8}  claimed that the usage of noble gas and high pressure may stabilize the nitrogen to form crown-shaped structures known as 18-crown-6-like-N$_8$. It was demonstrated that the 2D-N$_8$ layers persist to ambient pressure on decompression after the removal of xenon, comprising not only a layered bulk \textit{P}$\bar 1$ N$_8$ phase, but also a monolayer \textit{P}$\bar3$ N$_8$, which is dynamically and thermally stable. 
Lately, the  2D-N$_8$ compound was synthesized by employing a laser heating in a diamond anvil cell technique using a mixture of KN$_8$ and nitrogen as initial compounds\,\cite{sui}. Furthermore, it was also confirmed that the structure exhibits the 2D form.
Pitié et al.\,\cite{pitie} extended the first investigation by predicting the new structures of porous polynitrogen that accommodate metal ions.  The structure was designated as 2D-N$_8$, derived from the predicted high-pressure stable layered compound XeN$_8$ at 100\,GPa\,\cite{sui}. 
The resulting structures of a covalent nitrogen monolayer  2D-N$_8$ consist of a ring containing six N=N units and six three-coordinated nitrogen atoms in sp$^3$ hybridization. 

Motivated by the previous theoretical predictions and experimental realization, the present study aims to perform comprehensive calculations of the geometry, stability, and electronic properties calculation of the 2D-N$_8$ structure. The adsorption characteristics of metal ions commonly used in battery applications will be specifically examined.  Subsequently, the adsorption energies of the Li, Na, Mg, K, Ca, S, and Zn metals in the pristine 2D-N$_8$ structure will be investigated. The studied system's theoretical storage capacity (TSC) and open-circuit voltage (OCV) with adsorbed metals will be evaluated. Finally, the calculated migration energy barriers for the metals will be analyzed.

The paper is organized as follows. Sec.\,\ref{sec:comp} contains computational details and formulas used for all property predictions, including stability check, electronic properties, and evaluation of battery material performance. Sec.\,\ref{results} is dedicated to the results of properties analysis of the pristine 2D-N$_8$ and detailed evaluation of the adsorption characteristics of various metal ions. The last section is the conclusions, in which a summary of the results is presented.

\section{Computational details}
\label{sec:comp}
\subsection{Electronic structure calculation parameters}
Geometric and electronic properties of the pristine and boron-doped 18-crown-6-like-N$_8$ structures before and after the intercalation of the metal-ions were carried out by DFT calculations implemented in Vienna ab initio simulation package (VASP)\,\cite{40}, using the plane wave basis set with an energy cutoff of 520\,eV, the projected augmented wave potentials (PAW)\,\cite{41}. 
The generalized gradient approximation (GGA) with the Perdew, Burke, and Ernzerhof functional (PBE) is utilized for the exchange-correlation functional\,\cite{42}. K-mesh sampling used the Monkhorst-Pack 7:7:1 grid\,\cite{43}. The electronic iterations convergence was set to be 10$^{-5}$\,eV algorithm for energy, 0.01\,eV/\AA~for Hellmann–Feynman forces and real space projection operators. All structures were checked to be relaxed within this accuracy setting. 

\begin{figure}[h!]
\centering
 \includegraphics[width=\linewidth]{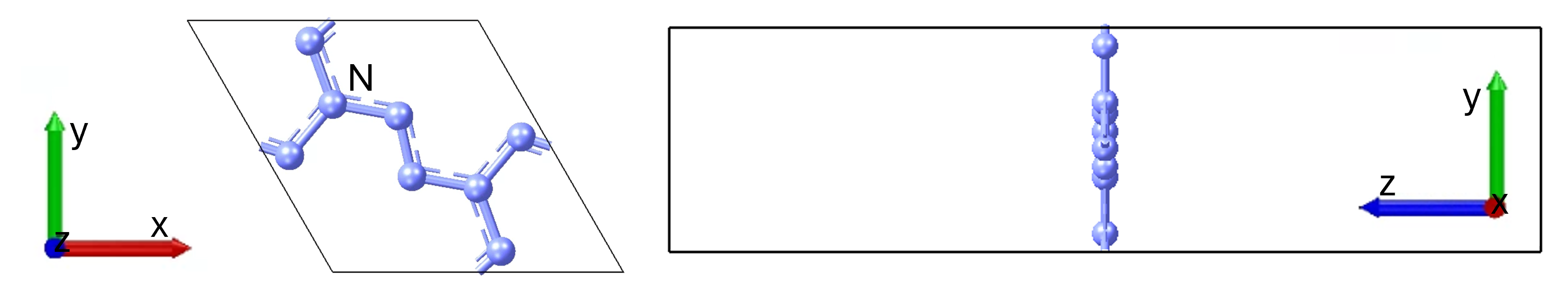}
\caption{Top and side view of the unit cell of the pristine 18-crown-6-like-N$_8$ structure monolayer used in calculations.}
\label{fig_cell}
\end{figure}

To better describe the long-range van der Waals interaction between monolayers and adsorbed metal ions, Grimme’s DFT-D$_3$ scheme was considered\,\cite{44}. To quantitatively assess the charge transfer that occurs between the metal ions and the electrode, the Bader charge method\,\cite{45} was employed. The lattice parameters of pristine N$_8$ monolayers’ unit cell, obtained after full relaxation, were found to be a = b = 5.925\,\AA. Periodic boundary conditions were applied in two dimensions, \textit{x} and \textit{y}, while a vacuum region longer than 20\,\AA~along the \textit{z}-direction was set to avoid interactions between adjacent layers, ensuring isolation of a single layer as indicated in the unit cell representation of Fig.\,\ref{fig_cell}. 

\subsection{Dynamic and thermal calculation parameters}
\label{sec:dynamic}
The kinetic process of diffusion pathways and the energy barrier was calculated using the nudged elastic band method of the climbing image (CI-NEB)\,\cite{48}, and the intermediate images were relaxed until the forces were less than 0.03\,eV/\AA. To explore the dynamic stability of the N$_8$ monolayer, the phonon dispersion spectrum was calculated using the finite displacement method implemented in VASP. Ab initio molecular dynamics simulations (AIMD) were performed within the NVT ensemble and the Nose-Goover thermostat with a time step of 2\,fs and a total simulation time of 20\,ps at 500\,K to evaluate the thermal stability.

\subsection{Mechanical properties calculation parameters and formulas}
\label{sec:mechanical}
Mechanical stability was evaluated by calculating the elastic constants. The hexagonal phase of the monolayer used in the calculations is characterized by three constants: C$_{11}$=C$_{22}$, which represent the material's in-plane stiffness under uniaxial strain applied in the \textit{x} or \textit{y} direction; C$_{12}$, which describes resistance to biaxial deformation; and C$_{66}$, which quantifies resistance to distortion caused by shear stress.   

Strains were applied to the material ranging from -2.5\,\% to 2.5\,\% in increments of 0.5\,\%. Then, elastic constants were estimated from the strain energy-strain curves according to the following formulas:
\begin{equation}
    \left.C_{11}  = \frac{1}{A_0} \left(\frac{\partial^2{U}}{\partial{s}^2}\right)\right\vert_{s=0},
\end{equation}
for uniaxial strain, and
\begin{equation}
    \left.2(C_{11}+C_{12})  = \frac{1}{A_0} \left(\frac{\partial^2{U}}{\partial{s}^2}\right)\right\vert_{s=0},
\end{equation}
for biaxial strain,
where $A_0$ is the equilibrium unit cell area, $U$ is the strain energy, $s$ is the applied strain. The last coefficient can be obtained as  
\begin{equation}
    C_{66} = \frac{C_{11} - C_{12}}{2}.
\end{equation}

Using calculated elastic constants, the Born-Huang stability criteria were verified using the following equations\,\cite{49}:
\begin{equation}
    C_{11}C_{12}  > 0,
    \label{eq:4}
\end{equation}
\begin{equation}
    C_{66} > 0,
    \label{eq:5}
\end{equation}
\begin{equation}
    C_{11}-C_{12} > 0.
    \label{eq:6}
\end{equation}

Additionally, elastic constants allow for estimation of the mechanical properties of the material, such as Young’s modulus in the plane: 
\begin{equation}
    Y = \frac{C_{11}^2 - C_{12}^2}{C_{11}^2},
\end{equation}
and Poisson’s ratio 
\begin{equation}
    \vartheta = \frac{C_{12}}{C_{11}}.
    \label{eq:example_label}
\end{equation}

Bulk modulus can be defined as the second derivative of the strain-energy curve:
\begin{equation}
    \left.G  = A_0 \left(\frac{\partial^2{U}}{\partial{A}^2}\right)\right\vert_{A = A_0}.
\end{equation}

\subsection{Battery performance characteristics' formulas}
\label{sec:battery_performance}
The averaged adsorption energy (E$_{ads}$) of metal-ion onto the pristine 2D-N$_8$ is calculated as follows:
\begin{equation}
E_{\text{ads}} = \left( E_{\text{metal-N$_8$}} - \left( E_{\text{bare}} + x \cdot E_{\text{metal}} \right) \right) / x.
\label{eq:adsorption}
\end{equation}
Here,  $E_{metal-N8}$ and $E_{metal}$ are the calculated energies of metal ions adsorbed on the N$_8$ monolayer and the energy of isolated metal ions, respectively, and \textit{x} is the number of metal ions adsorbed by the monolayer.  

To check the metal-ion-adsorbed thermodynamic stability the idea of a convex hull is convenient to use. For the case of a system with added ions and the fixed composition of host material, the following equation can be consequently obtained\,\cite{mbenes,wu}:
\begin{equation}
E_{\text{form}} = \left( E_{\text{metal-N$_8$}} - \left( E_{\text{bare}} + x \cdot E_{\text{metal}} \right) \right) / (x+1).
\label{eq:formation}
\end{equation}

The OCV value can be determined using the Gibbs free energy change ($G_{cell}$) of the half-cell reaction according to the following equation:
\begin{equation}
    OCV = -\frac{\Delta G_{cell}}{zF},
\end{equation}
where \textit{F} is the Faraday constant and \textit{z} is the number of valence electrons during the metal-ion adsorption process. Changes in entropy and volume during a half-cell reaction are often assumed to be minimal and insignificant\,\cite{50}. Therefore, the open-circuit voltage is calculated as the following equation:
\begin{equation}
  \mathrm{OCV} = -\frac{E_{\text{ads}}}{zF},
\label{eq:ocv}
\end{equation}
where \textit{z} is the number of valence electrons (z = 1 for Li, Na and K, z = 2 for Ca and Mg). 
The following equation can resolve theoretical storage capacity:
\begin{equation}
C = \frac{z x_{max} F}{M_{N_8}},
\label{eq:tsc}
\end{equation}
where \textit{z} is the number of valence electrons, $x_{max}$ is the maximum number of adsorbed metal-ions, \textit{F} is Faradey constant (26801\,mA/mol) and $M_{N_8}$ is the relative atomic mass of 2D-N$_8$.

\section{Results and Discussion}
\label{results}
\subsection{Pristine 2D-N$_8$}
\label{pristine}
\subsubsection{Stability analysis}
\label{sec:stability_analysis}
The stability of model materials refers to the analysis and prediction of how materials behave under various conditions. It allows us to predict how a material will perform in real-world situations, such as extreme temperatures, pressures, or chemical environments, without physically testing it in all scenarios. Here are some key aspects of the stability of the modeled materials: thermal, kinetic, mechanic, thermodynamic, and electronic. 

The \textbf{thermal stability} of the system has been checked by AIMD simulations at 500\,K, and the variation of the potential energy during the simulation is presented in Fig.\,\ref{fig_cell}\,a. The N$_8$ monolayer demonstrates good thermal stability. It maintains the initial configurations after 10\,ps with holes similar to the equilibrium structure and preserving the bond lengths.
\begin{figure}[h!]
\centering
\begin{minipage}{0.55\textwidth}
\includegraphics[width=\linewidth]{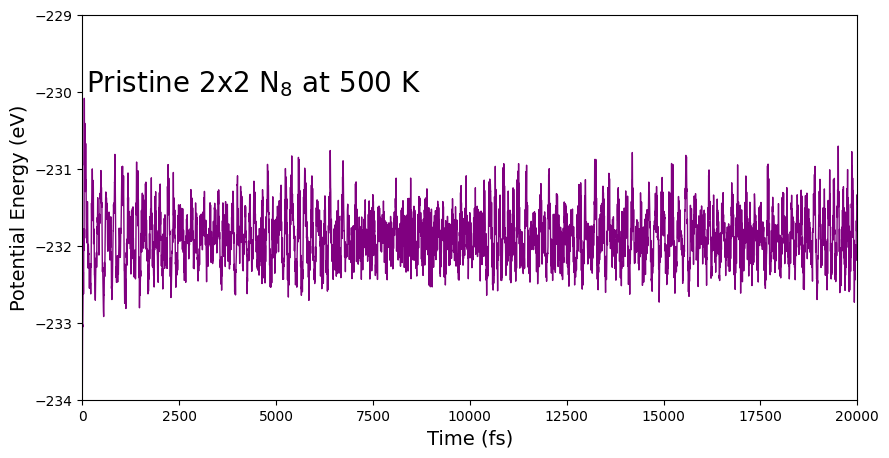} \\ (a)
\end{minipage}
\begin{minipage}{0.43\textwidth}
\includegraphics[width=\linewidth]{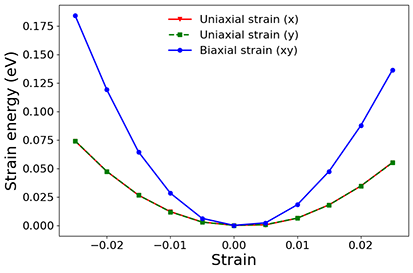} \\(b)
\end{minipage}
\caption{(a) AIMD at 500\,K for 2$\times$2-cell of N$_8$ monolayer, (b) variation of the total energy with applied uniaxial strain along the \textit{x}-axis, and \textit{y}-axis, and the biaxial strain along the \textit{x} and \textit{y}-axis.}
\label{fig_stability}
\end{figure}

The \textbf{mechanical stability} is essential in designing high-capacity cathode materials. Mechanical stability focuses on how a material responds to physical stresses (e.g., tension, compression, or shear). Models assess whether a material will deform, fracture, or retain its structural integrity under these forces. The pristine N$_8$ monolayer satisfies the mechanical stability criteria of Eqs.\,\eqref{eq:4}-\eqref{eq:6}. The stress-strain curves shown in Fig.\,\ref{fig_stability}\,b, were computed to estimate the elastic constants and mechanical properties.  The evaluated elastic constants are C$_{11} =$ C$_{22} =$ 621.7\,GPa,  C$_{12} =$ 145.4\,GPa, C$_{66} =$ 238.1\,GPa. The calculated Young modulus is 587.7\,GPa, Poisson’s ratio is 0.23, and the bulk modulus is 154.6\,GPa. 

Phonon spectra show the \textbf{dynamical stability} of the system. Namely, if any calculated phonon modes exhibit imaginary frequencies, it indicates that the structure is dynamically unstable. This typically might point out to distortions or phase transitions. The calculated phonon dispersions for 2D-N$_8$ monolayer in the structure depicted in Fig.\,\ref{fig_cell} possess imaginary phonon frequencies. This negative phonon mode is associated with the translational movement of the whole monolayer in the \textit{z}-direction. Reduced dimensionality often heightens the effect of long-range interactions, making the system more susceptible to instabilities. Consequently, in 2D systems, these interactions can cause the phonon modes responsible for out-of-plane displacements to become less stable, leading to negative\,\cite{phonon}.
To find a dynamically stable configuration, we first constructed a layered bulk structure with a reduced plane-to-plane distance between monolayers and fully optimized it.
The most energetically stable configuration was found to be in the form of twisted monolayers. This structure was confirmed to be dynamically stable with no imaginary frequencies. After that, three monolayers were surrounded by thick vacuum space as shown in SFig.\,1 of Supplementary.
In this case, the optimized distance between monolayers is 3.5\,\AA. The corresponding phonon dispersion curves are presented in SFig.\,2\,a, where positive frequencies indicate favorable dynamical stability. 
We also calculated the binding energy required to exfoliate one monolayer of N$_8$ from the constructed bulk material. The resulting binding energy, E$_{binding}$, is minimal, equal to 0.04\,eV/atom, demonstrating that the monolayer can easily be formed.
In addition, we also present the phonon curves for a single N$_8$ monolayer with an adsorbed metal ion, which, as we see in SFig.\,2\,b, stabilize the studied monolayer.

\begin{figure}[h!] 
\centering
\begin{minipage}{0.49\textwidth}
\includegraphics[width=\linewidth]{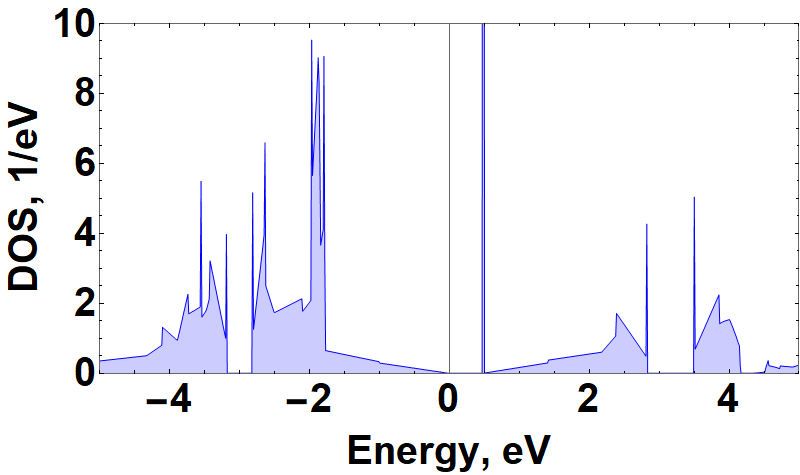} \\(a)
\end{minipage}
\begin{minipage}{0.49\textwidth}
\includegraphics[width=\linewidth]{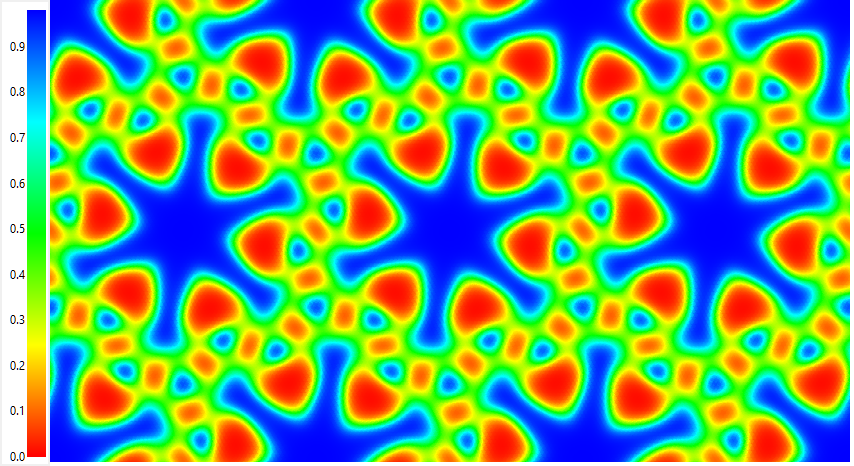}  \\ (b)
\end{minipage}
\caption{(a) Density of states and (b) ELF plot of pristine N$_8$ monolayer (from Fig.\,\ref{fig_cell}).}
\label{fig_electronic}
\end{figure}

In semiconductors and other electronic materials, the \textbf{stability of electronic structures} is crucial. If the material's band structure or electron configuration is unstable, it may not function correctly in electronic devices. Small-size band-gap semiconductors are promising materials for metal-ion battery cathodes. Moreover, metallic materials are desirable for both anode and cathode materials because of the reduced risk of possible overheating of electrodes. The analysis of the density of states (Fig.\,\ref{fig_electronic} a) shows that the system is a semiconductor with a band gap of 0.540\,eV in agreement with previously published results reporting 0.502\,eV\,\cite{pitie}. 
The structure of the band (and the corresponding DOS) of the N$_8$ system revealed an important feature of a flat band located right above the Fermi level, as seen as a sharp peak in the DOS. These bands in electronic systems are of significant interest because they imply unique and potentially transformative electronic properties. In a flat band, the energy of the electronic states is nearly constant, which means that the electrons in these states have almost no kinetic energy. This results in a high density of electronic states at the same energy, which can enhance interactions between electrons, such as those leading to superconductivity, magnetism, or other correlated phenomena. That might be the subject of further investigation.

The electron localization function (ELF) (Fig.\,\ref{fig_electronic}\,b) was calculated to describe electron localization in the material quantitatively. ELF values range from 0 to 1, where values close to 1 indicate strong electron localization (e.g., in covalent bonds), values near 0 correspond to delocalized electrons (e.g., in metals), and 0.5 represents the electron-gas state. The following color coding is commonly used in ELF visualization. The high ELF values between the nitrogen atoms in the 2D-N$_8$ layers indicate strong covalent bonds that stabilize its 2D structure.

\textbf{Thermodynamic stability} describes whether a material is in its most stable state, which means that it has the lowest possible energy configuration. Computational models can help predict whether a material will remain stable or whether it might undergo a phase change or decomposition over time. N$_8$ adsorption characteristics of Li, K, Ca, Na, Mg, S, and Zn were tested. To assess thermodynamic stability, the formation energy using Eq.\,\eqref{eq:formation} was calculated and plotted in SFig.\,3. The formation energies versus the metal-ion content form a convex hull that connects the lowest formation energies. The structures in the convex hull represent stable configurations. It is seen that the N$_8$ monolayer forms stable configurations with considered metal-ions. 

\begin{figure}[h!]
\centering
\includegraphics[width=\linewidth]{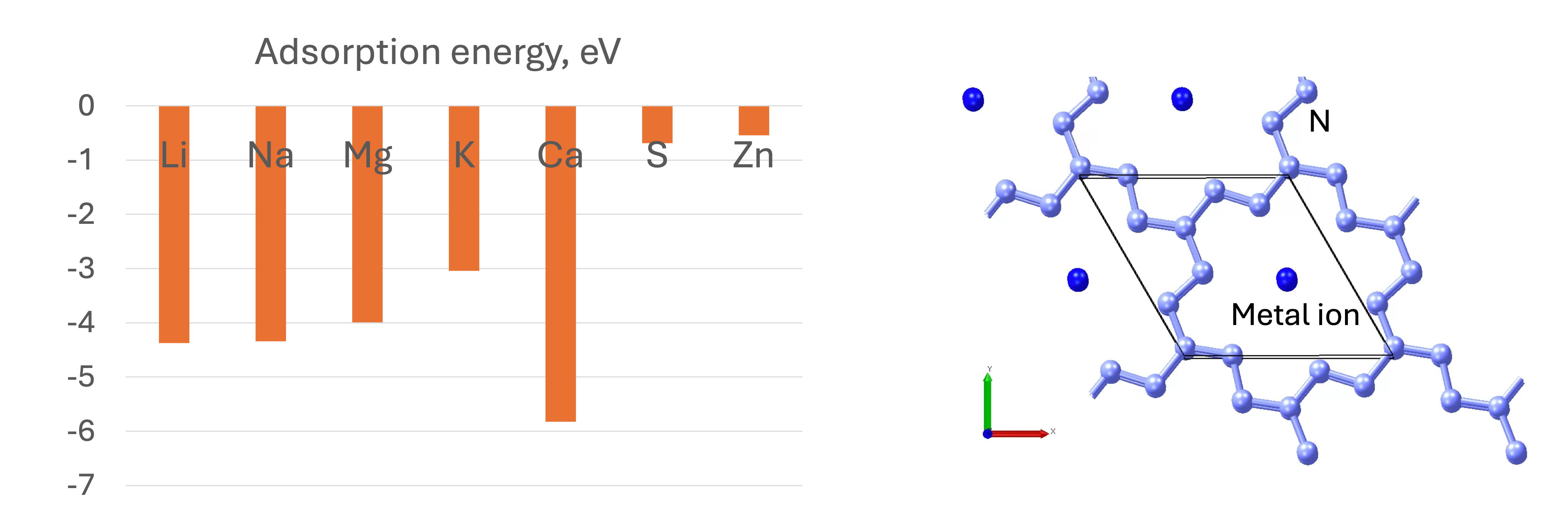}
\caption{Comparison of adsorption energies of a single Li/K/Ca/Na/Mg/S/Zn ion on the surface of the pristine 2D-N$_8$ monolayer with denoted favorable metal-ion position.}
\label{fig_comparison}
\end{figure}

As presented in Fig.\,\ref{fig_comparison} the adsorption energies are well below zero for most of the metal ions tested with the lowest and consequently the most stable configurations formed with Ca ions. The most probable adsorption position (for the concentration of one impurity ion per unit cell) was found to be in the middle of the N-ring, as indicated in Fig.\,\ref{fig_comparison}. The exact positions at the maximum possible concentration for selected metal ions will be described in detail in the following sections.

The presented analysis revealed the stability and good conducting characteristics of the N$_8$ monolayer system, confirming its possible application. It should be pointed out that the stability was also confirmed by experimental realization as was proven in Ref.\,\cite{sui}, where the single monolayers were also found to be stable. 
The next section will evaluate the metal-ions' adsorption features and battery performance characteristics, forming stable configurations.

\subsubsection{Adsorption of Li/Na/Mg/K/Ca metal ions}
\label{pristine_ads}
In the present section, the metal ions with the lowest adsorption energies, namely Li/Na/Mg/K/Ca, will be analyzed by investigating the N$_8$ monolayer structure with the highest stable number of metal-ions adsorbed on one side of the surface. 
In the next stage, the electrochemical efficiency of the 2D electrode material will be analyzed by carefully calculating the adsorption energies for an increasing number of metal ions on the surface of the electrode material. An important aspect is that only one side of metal-ions adsorption was presented here. 
Each 2D-N$_8$ monolayer with adsorbed metal ions will be tested for thermal stability by AIMD simulation at 500\,K.  

\begin{figure}[h!]
\centering
\includegraphics[width=\linewidth]{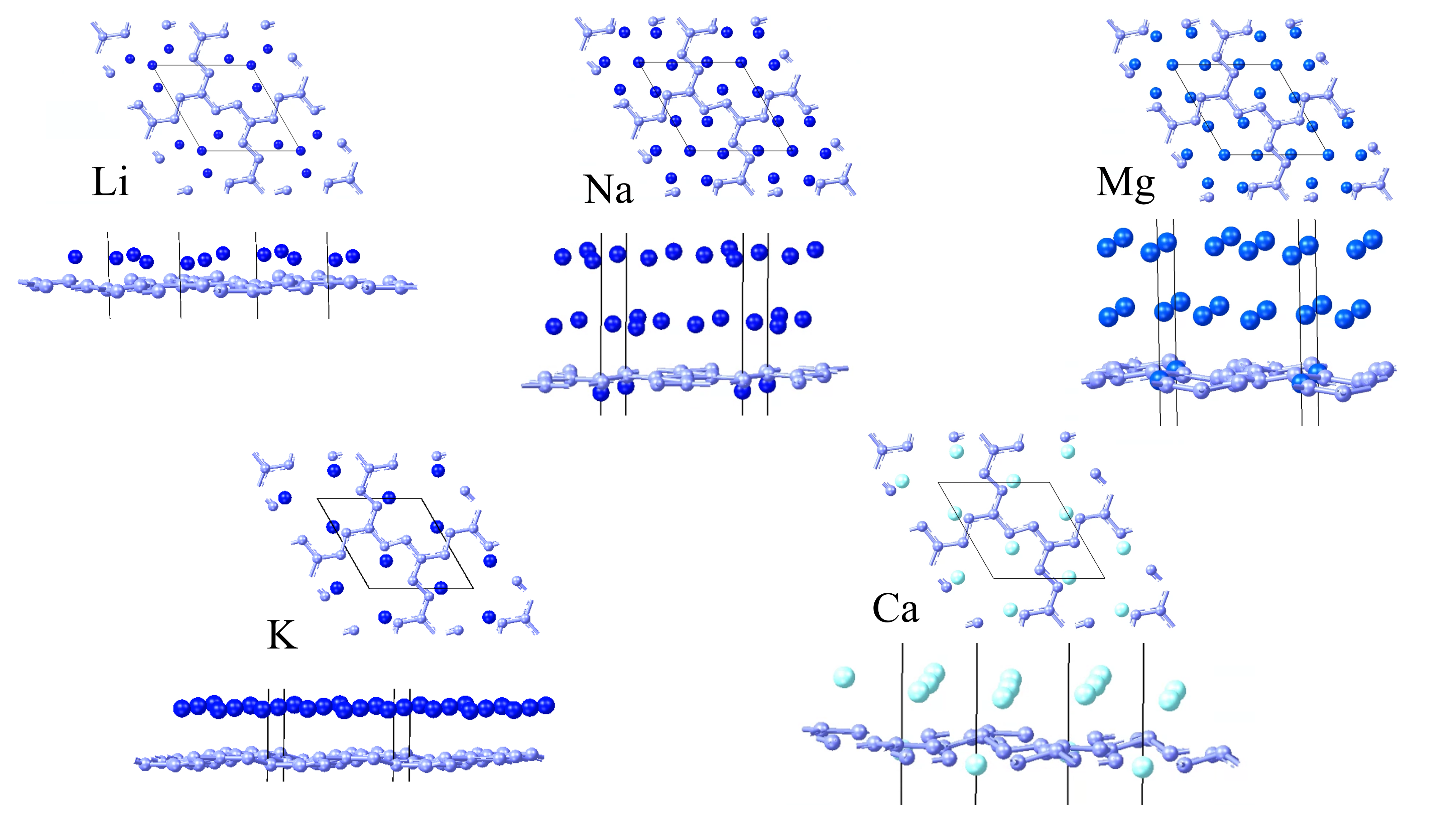}
\caption{Top and side view of pristine 2D-N$_8$ with the maximum amount of Li/Na/Mg/K/Ca metal ions on the surface at the optimized positions.}
\label{fig_cells}
\end{figure}

Fig.\,\ref{fig_cells} collects the top and side view of the 2D-N$_8$ surface with the maximum number of corresponding metal-ions. As seen from the structures, the planar structure is preserved for all cases with minor buckling. 

For the most abundant \textbf{Li metal ion} the maximum filling of the unit cell corresponds to the situation with 4 Li ions, which are located above the monolayer by 1.13-1.17\,\AA~with a negligible buckling of the distribution of Li ions on the surface of the planar monolayer. The position depends on the number of adsorbed ions. In particular, one Li-ion per unit cell occupies the middle of the pores, whereas the increased ions are distributed above the surface homogeneously. 
\textbf{Na-ions} adsorption was analyzed in the same way. The main difference from Li is that the number of adsorbed ions per unit cell increases significantly as presented in Fig.\,\ref{fig_cells}. We see that 7 K-ions homogeneously occupy the pore between N-N bonds, forming a few layers that are located equidistant above the monolayer. At the same time, one ion occupies the position within the surface slightly below it. 
A similar feature was observed for \textbf{Mg-ions} when up to 7 ions might be effectively adsorbed. 
Structure analysis (Fig.\,\ref{fig_cells}) reveals that \textbf{K-ions} are located above the monolayer by 2.17-2.46\,\AA, which is higher than for Li. At the same time, during adsorption, 2D-N$_8$ becomes slightly wavy (Fig.\,\ref{fig_cells}).
The last considered metal-ion is \textbf{Ca-ion}. It has a similar filling of energy evolution with a maximum of 3 adsorbed ions. The planar structure distortions are most prominent for that type of adsorption.

All investigated metal-ions form thermodynamically stable structures, as confirmed by AIMD simulations at 500\,K, as presented in SFig.\,4. The energy evolution in time shows that the structure remains planar, preserving the initial crown-type structure.

An additional aspect of the stability of the 2D-N$_8$ monolayer is that the presence of metal ions on the monolayer leads to the dynamic stabilization of the system. As was discussed in Sec.\,\ref{sec:stability_analysis} the monolayer N$_8$ surrounded with a vacuum higher than 3.16\,\AA~leads to the occurrence of imaginary frequencies. However, no negative modes are observed when metals are adsorbed (SFig.\,5).

\subsubsection{Adsorption of Li/Na/Mg/K/Ca metal ions and battery performance characteristics}
\label{pristine_adsorption}
The electrochemical efficiency of electrode material in metal-ion batteries is mainly characterized by strong metal bonding on the material surface. In other words, the host material must exhibit strong adsorption and correspondingly large negative adsorption energy.

\begin{figure}[h!]
\centering
\begin{minipage}{0.32\textwidth}
\includegraphics[width=\linewidth]{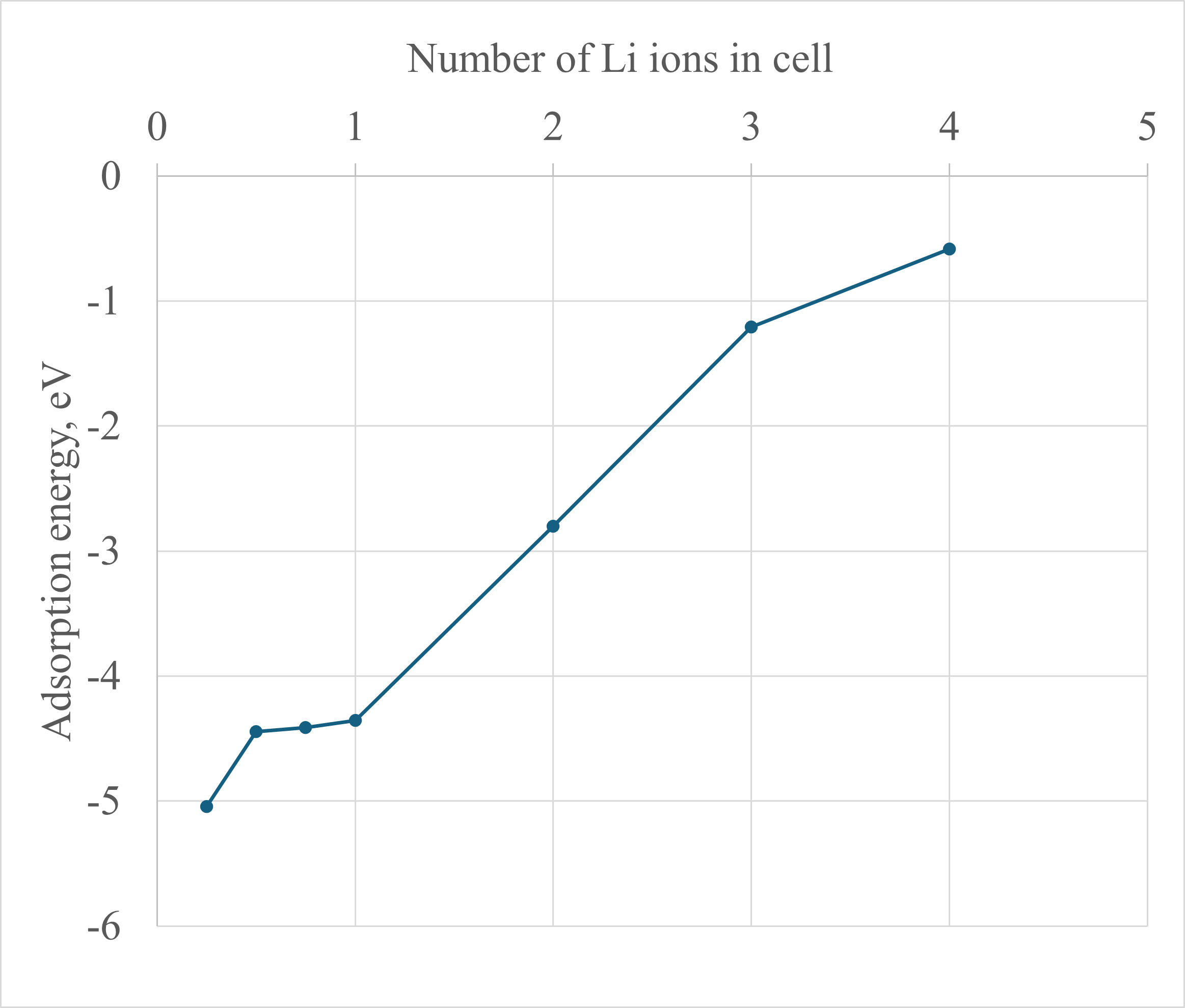}  
\end{minipage}
\begin{minipage}{0.32\textwidth}
\includegraphics[width=\linewidth]{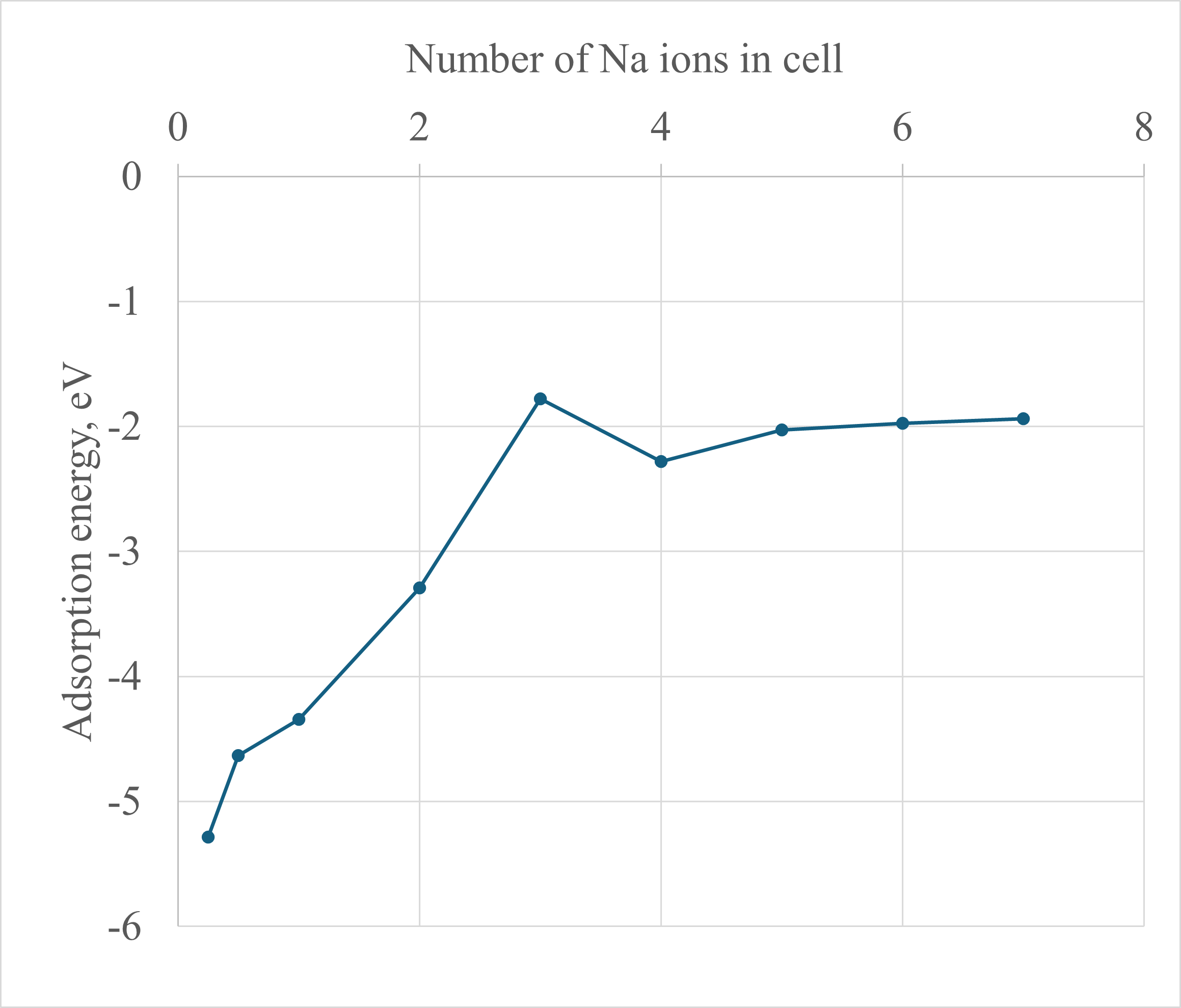}
\end{minipage}
\begin{minipage}{0.32\textwidth}
\includegraphics[width=\linewidth]{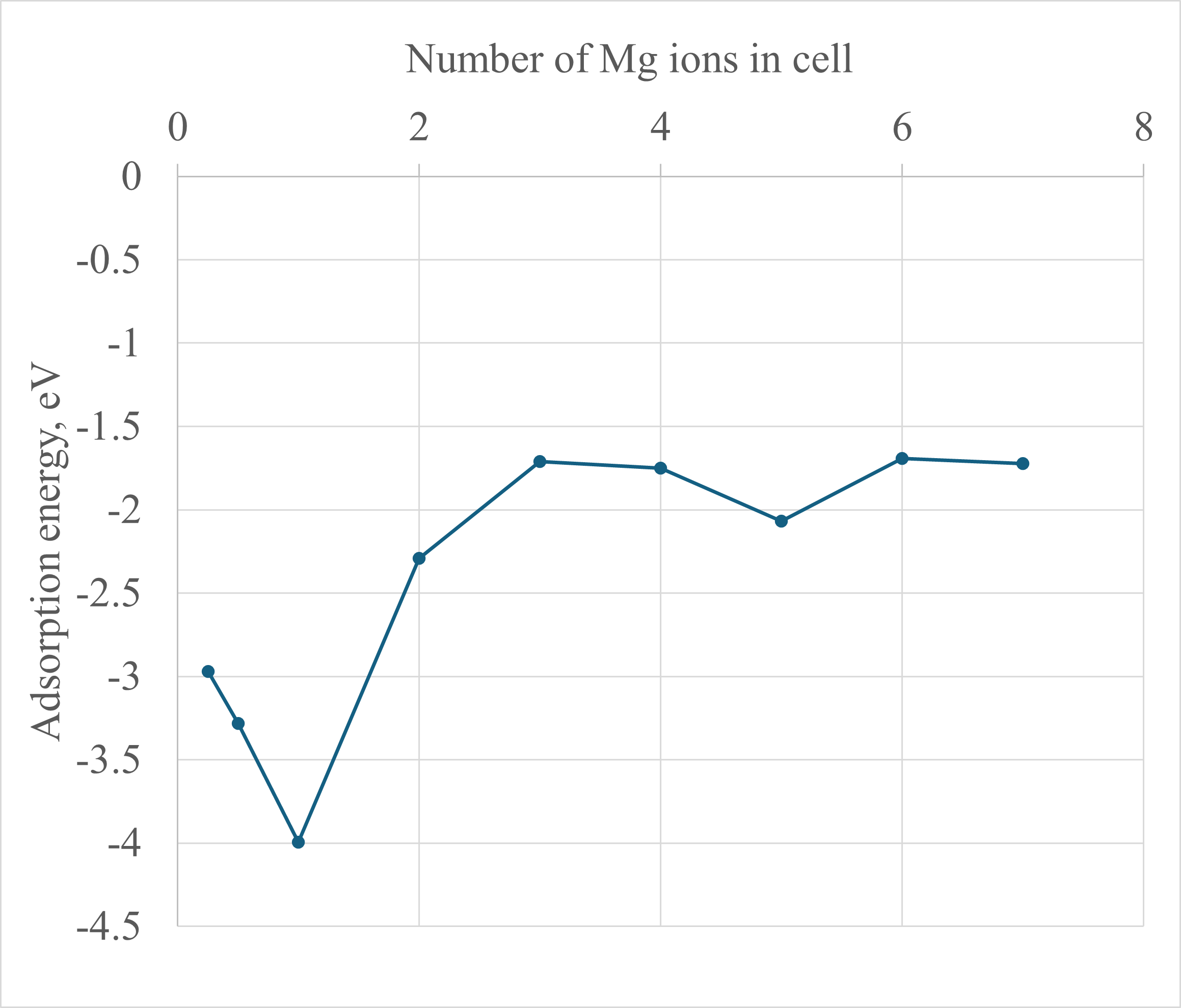}
\end{minipage}
\begin{minipage}{0.32\textwidth}
\includegraphics[width=\linewidth]{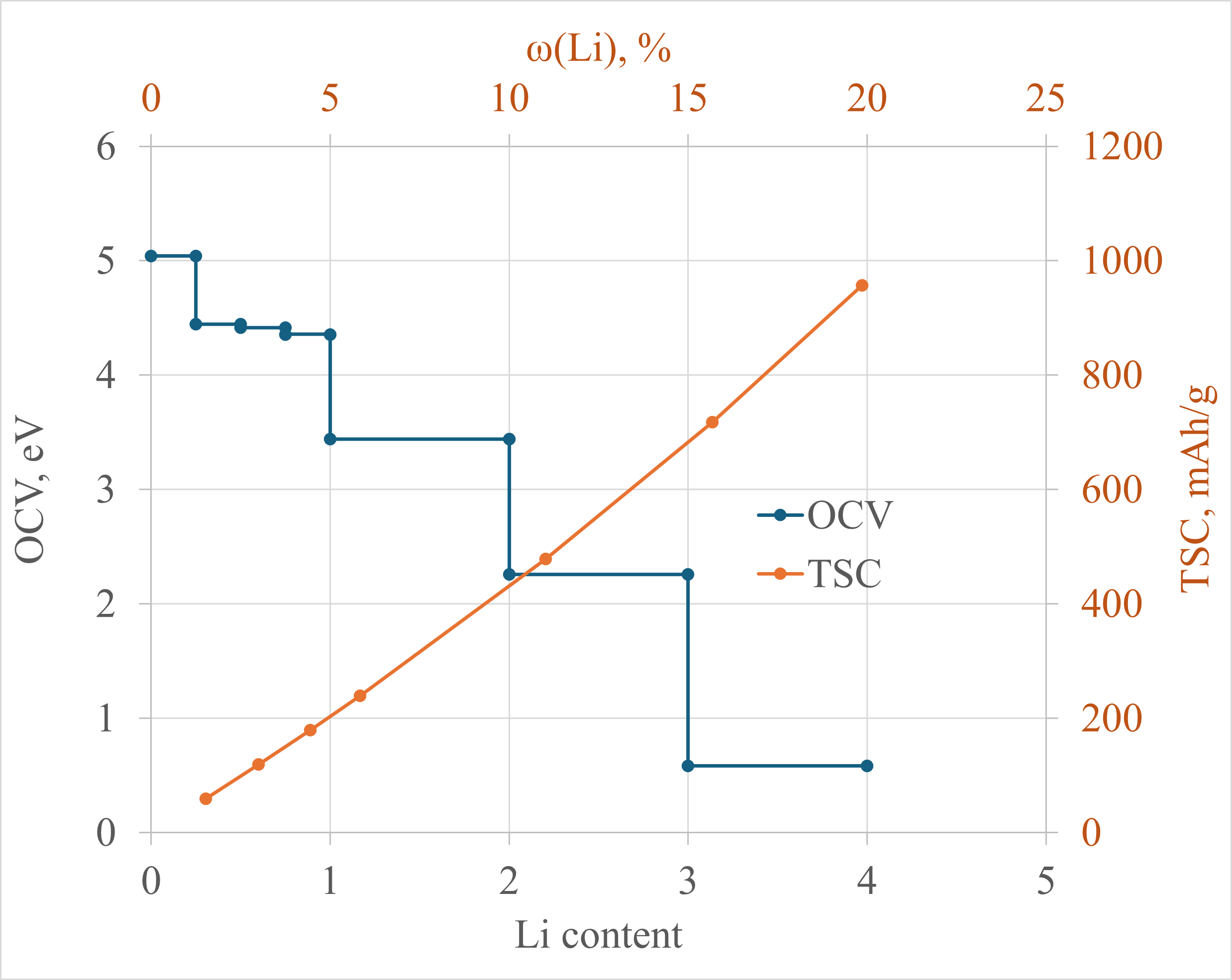} \\ \centering Li
\end{minipage}
\begin{minipage}{0.32\textwidth}
\includegraphics[width=\linewidth]{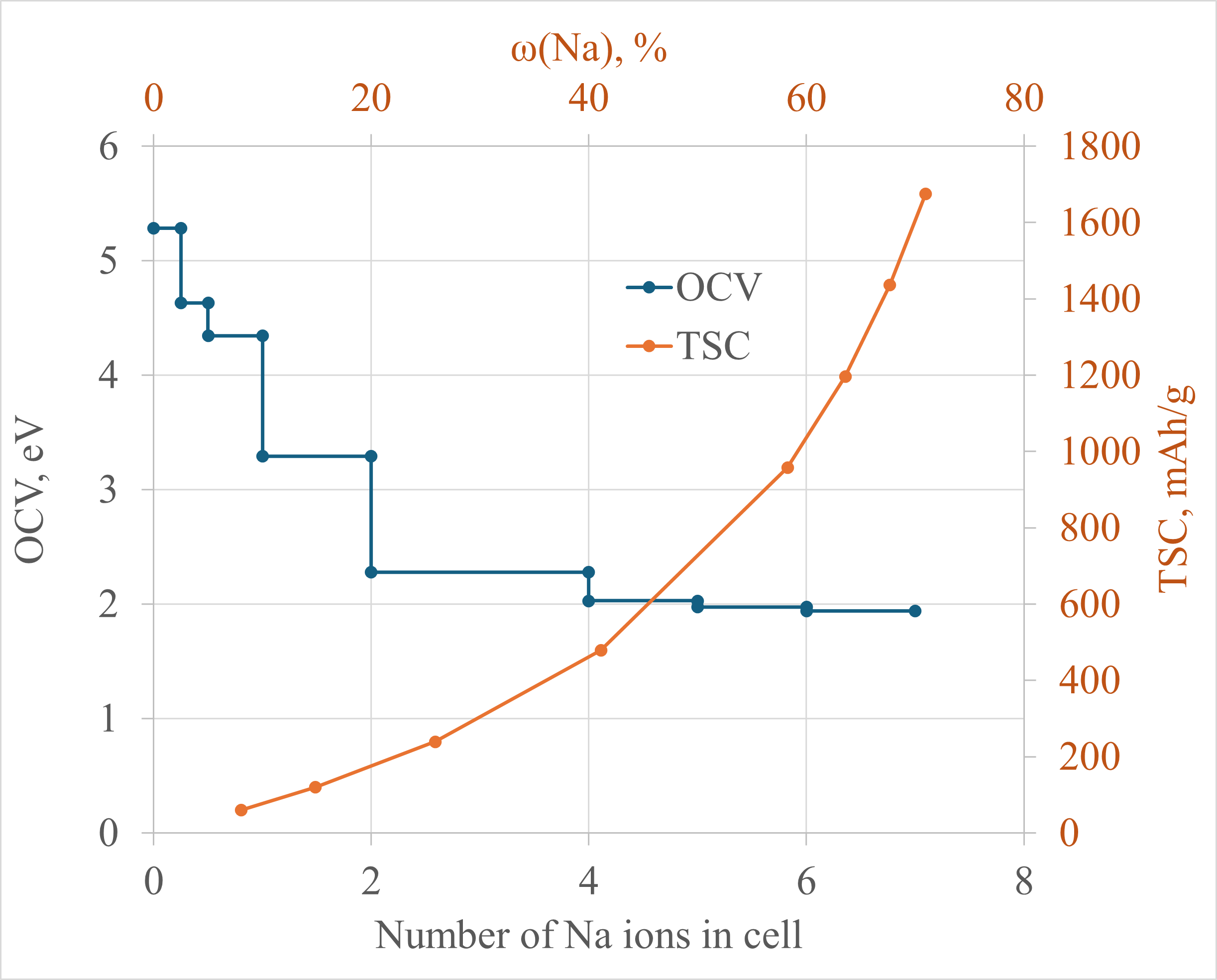} \\ \centering Na
\end{minipage}
\begin{minipage}{0.32\textwidth}
\includegraphics[width=\linewidth]{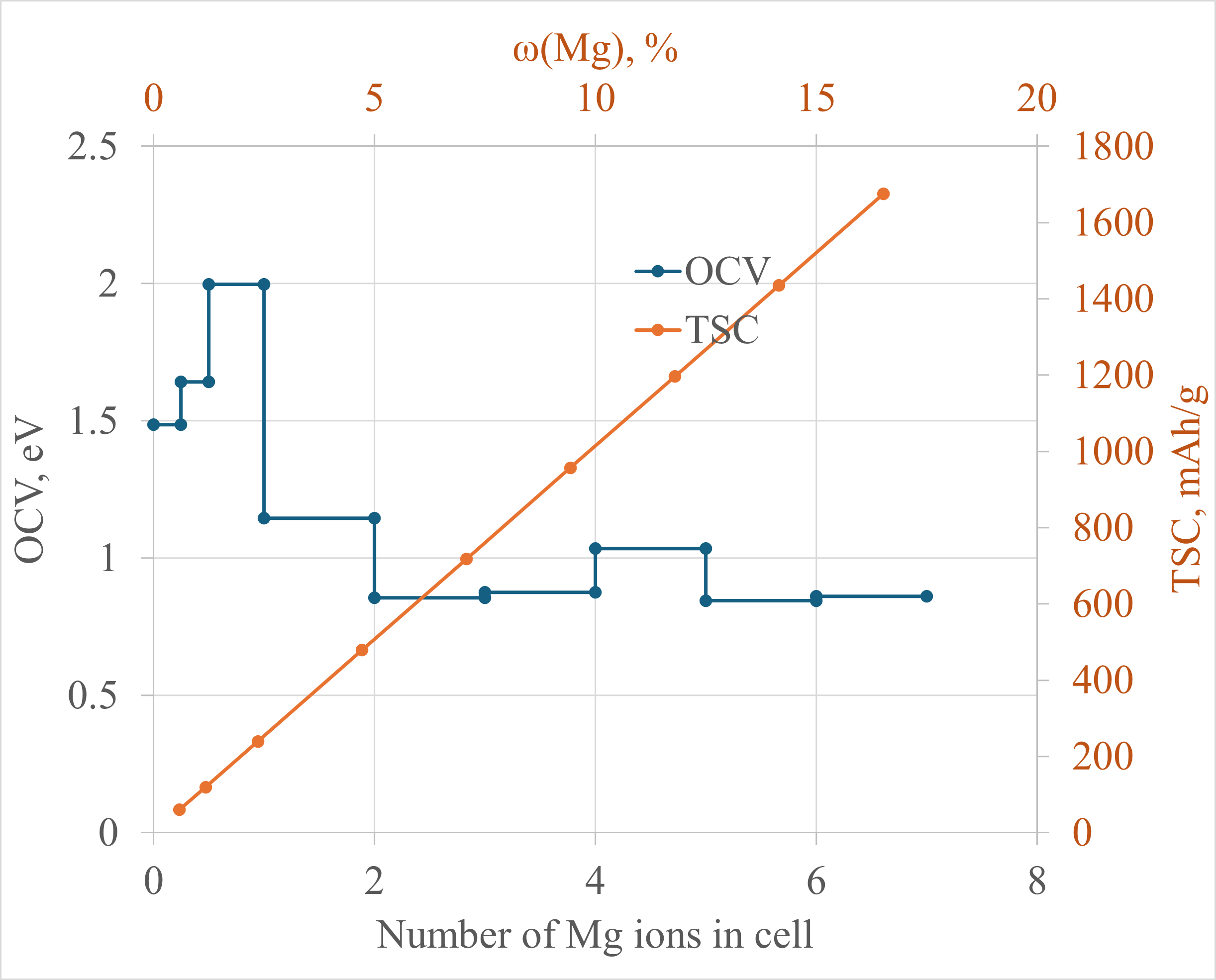} \\ \centering Mg
\end{minipage} \\
\begin{minipage}{0.32\textwidth}
\includegraphics[width=\linewidth]{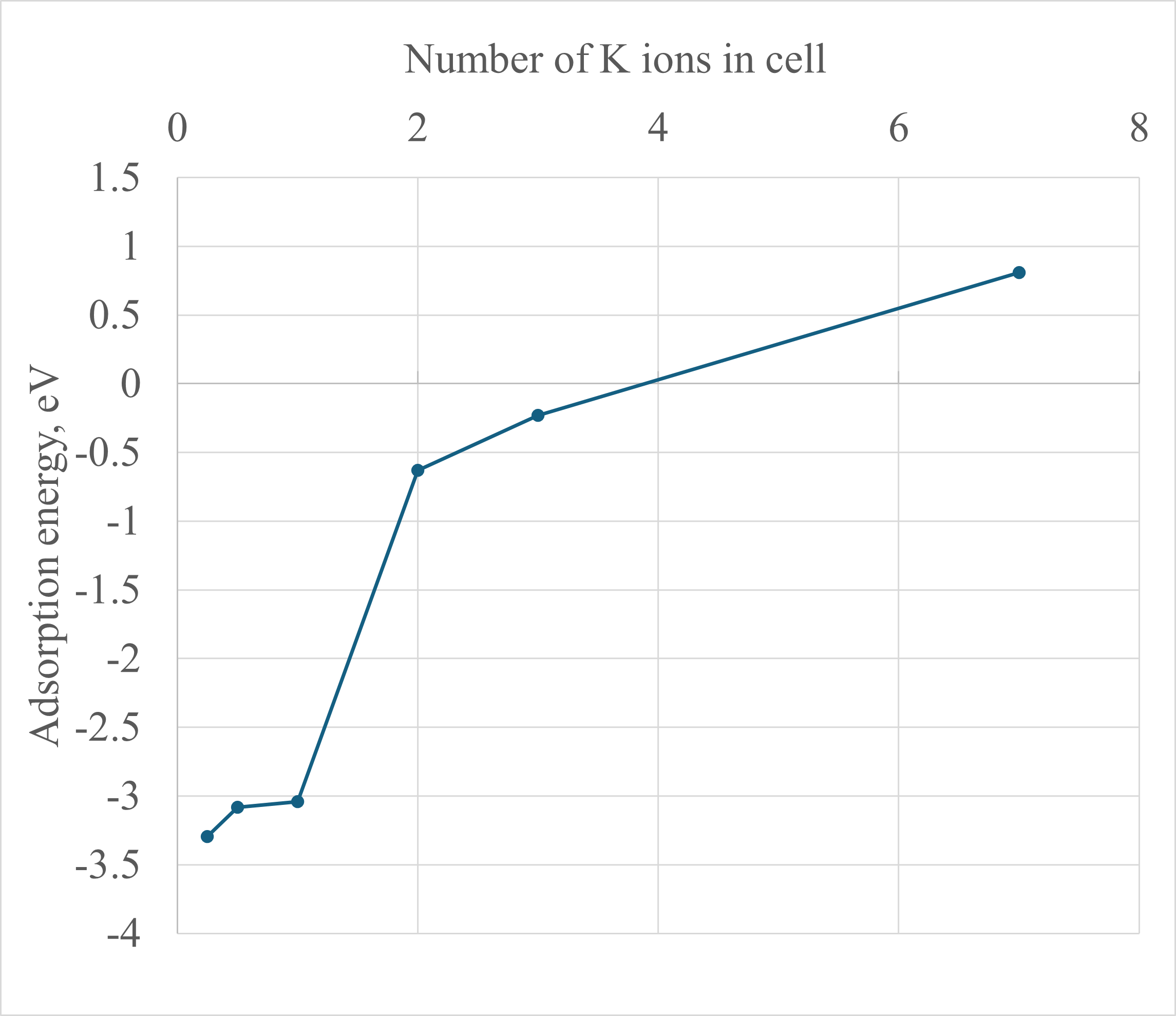} 
\end{minipage}
\begin{minipage}{0.32\textwidth}
\includegraphics[width=\linewidth]{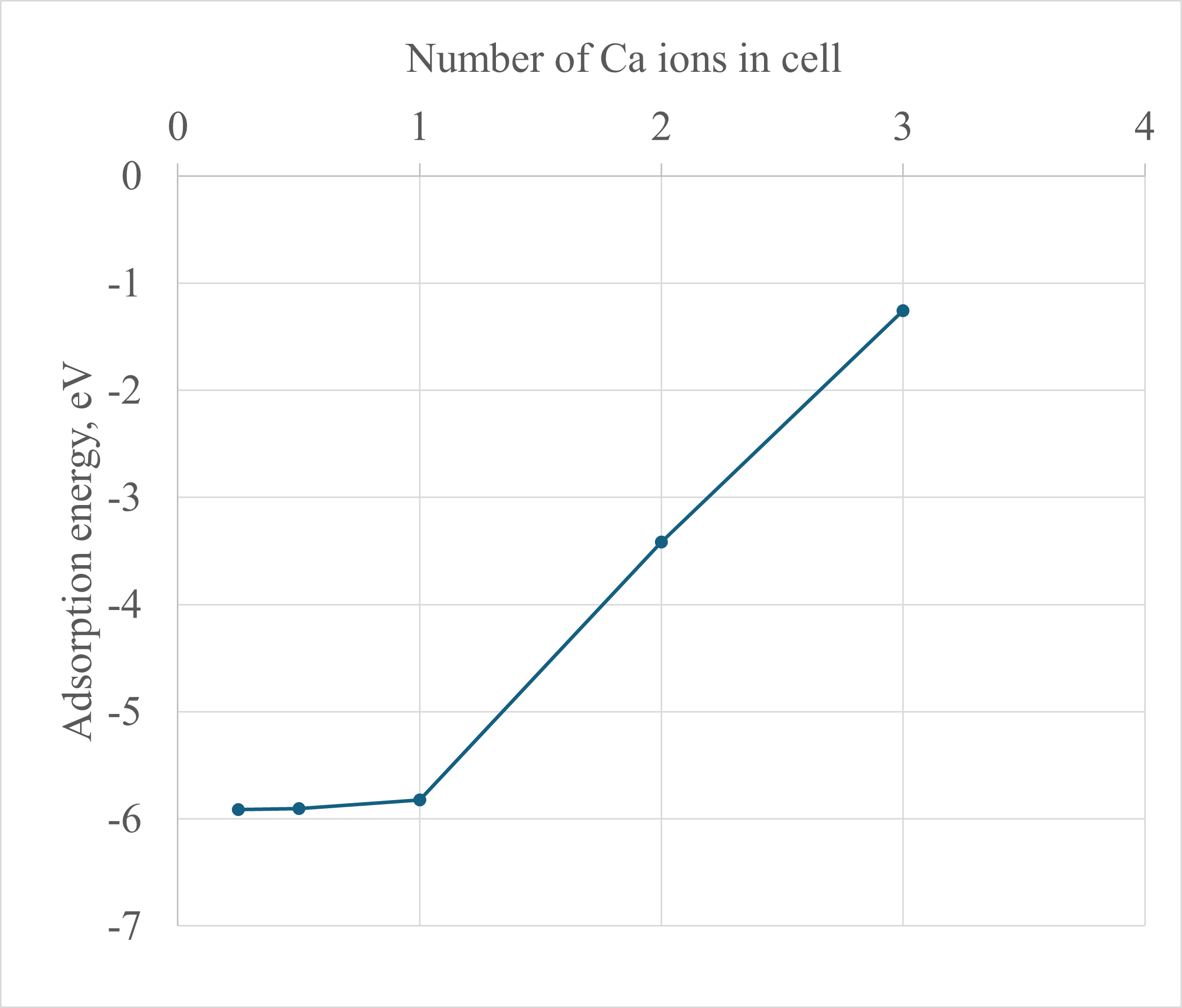}
\end{minipage} \\
\begin{minipage}{0.32\textwidth}
\includegraphics[width=\linewidth]{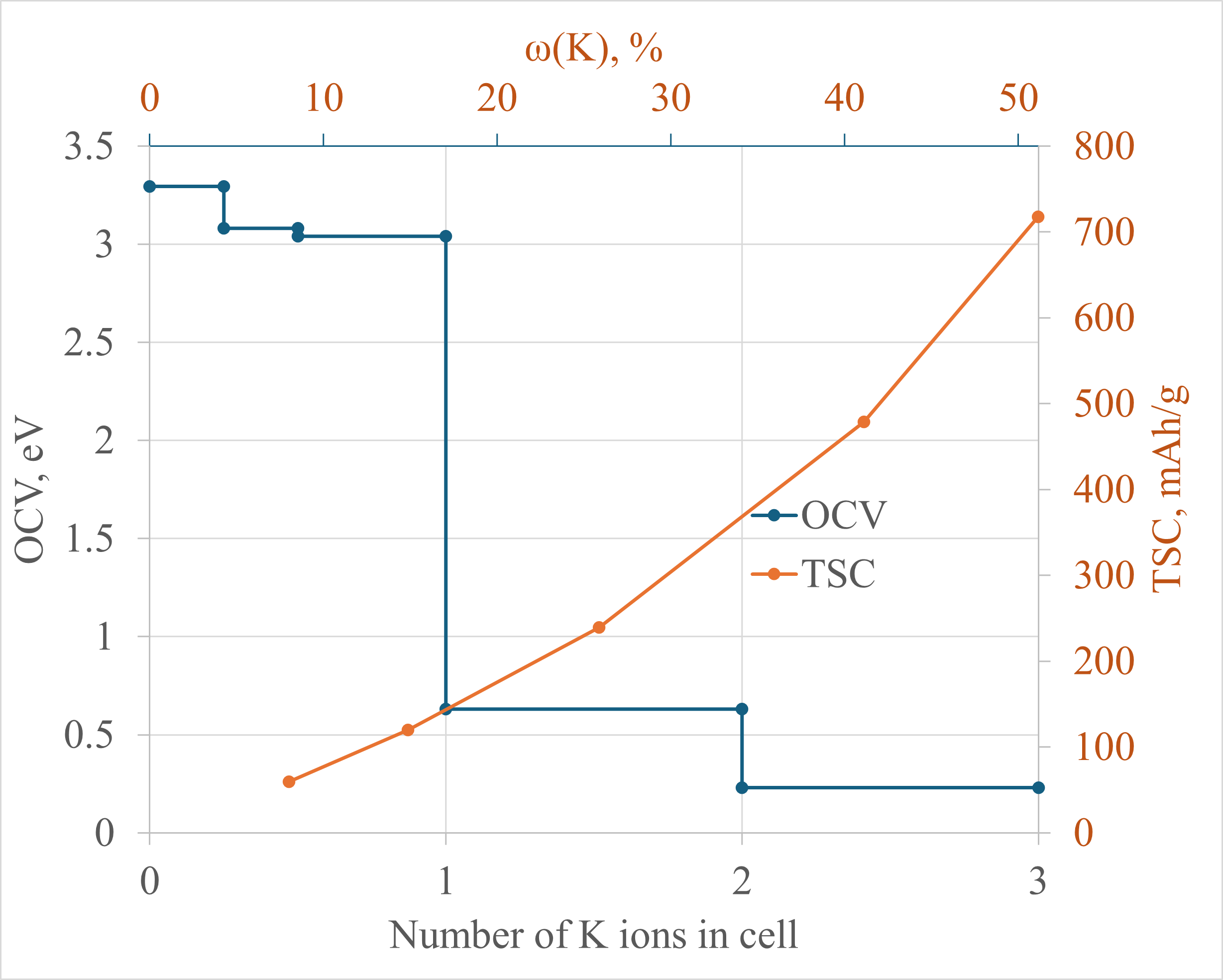} \\ \centering K
\end{minipage}
\begin{minipage}{0.32\textwidth}
\includegraphics[width=\linewidth]{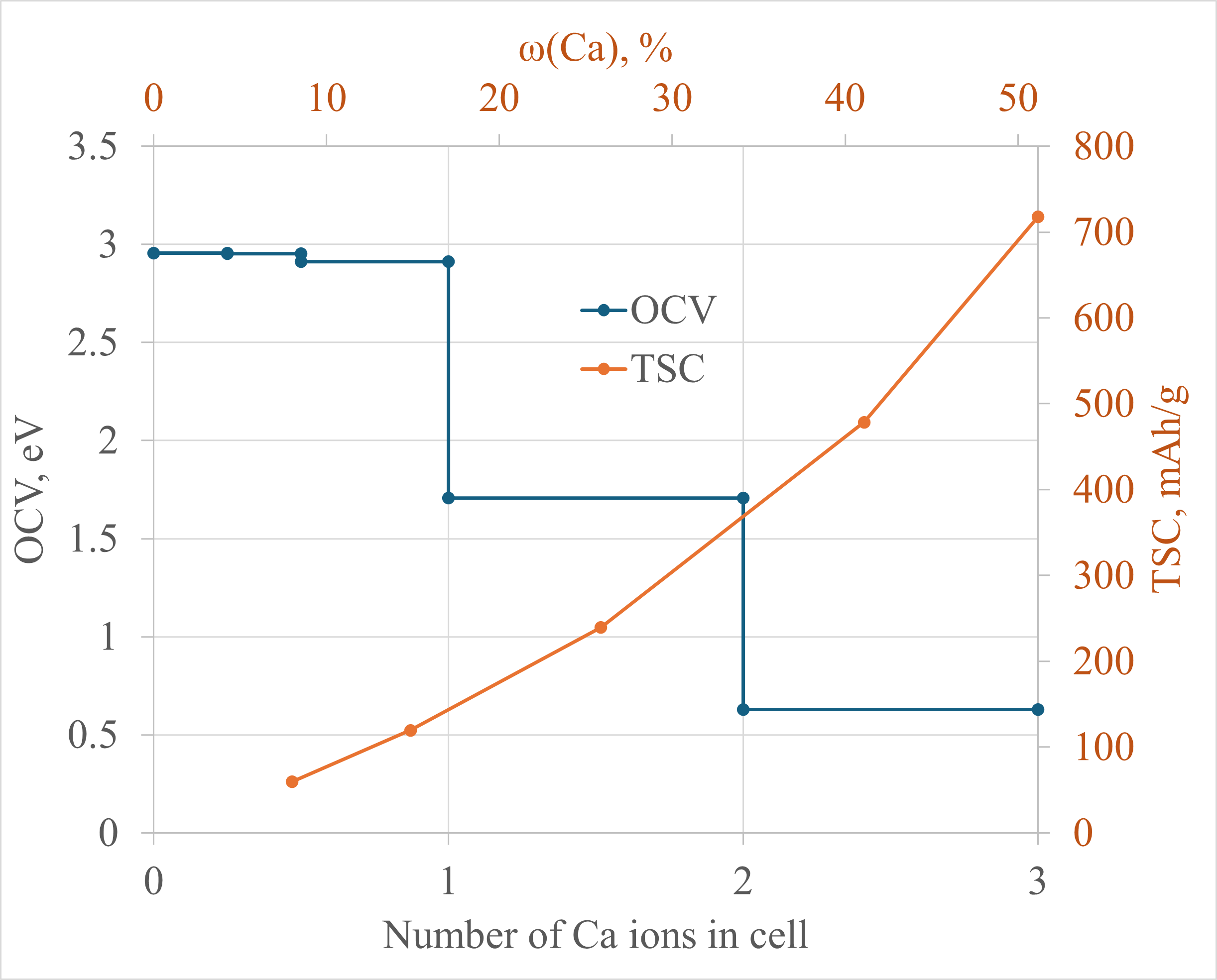} \\ \centering Ca
\end{minipage}
\caption{Adsorption energies of metal-ions and open circuit voltage (OCV) and theoretical storage capacity (TSC) of K-ions on the pristine 2D-N$_8$ monolayer.} 
\label{fig_adsorption}
\end{figure}

The adsorption energies were calculated according to Eq.\,\eqref{eq:adsorption} and plotted in Fig.\,\ref{fig_adsorption} for all the ions considered. We present the only regions with negative adsorption energies and stable configurations achieved during optimization. The situations depicted in Fig.\,\ref{fig_cells} correspond to the rightmost point on the graphs.
We obtained that the pristine N$_8$ monolayer can form a stable configuration with negative adsorption energy with a maximum of 4/7/7/7/3 ions of Li/Na/Mg/K/Ca adsorbed on one side. Interestingly, the formation energy curve for Na and Mg ends up with a plateau when a higher amount of adsorbed ions on one side with negative energy is hypothetically possible.

The increased number of adsorbed ions leads to unstable structures, which can be overcome by placing the ions on the other side of the 2D-N$_8$ monolayer. That type of adsorption is also thermodynamically stable. Moreover, the adsorption energy comes to a plateau for an amount of metal-ions larger than 4.
However, such adsorption leads to prominent distortion of the monolayer associated with out-of-plane buckling and, consequently, to non-planar structures.

An important aspect here is that the adsorption energy calculations were also performed for the three-layered structure surrounded by a vacuum slab. In Sec.\,\ref{sec:stability_analysis} such structure geometry was shown to be more dynamically stable than a single monolayer. The calculated adsorption energy for both single monolayer and three-layered structures are the same.

The OCV and TCS calculated by Eqs.\,\eqref{eq:ocv} and \eqref{eq:tsc} are presented in Fig.\,\ref{fig_adsorption}. The highest TSCs calculated for the Na and Mg ions are 1675\,mAh/g. These values are much higher than for graphene (372\,mAh/g\,\cite{endo}). The OCV range is an important characteristic since it indicates the potential of the cathode material to store metal-ions during discharging. The high mean OCV values for all metal-ions were found, which is desirable for cathode materials~\cite{fen}. 

\begin{figure}[h!]
\centering
\begin{minipage}{0.55\textwidth}
\includegraphics[width=\linewidth]{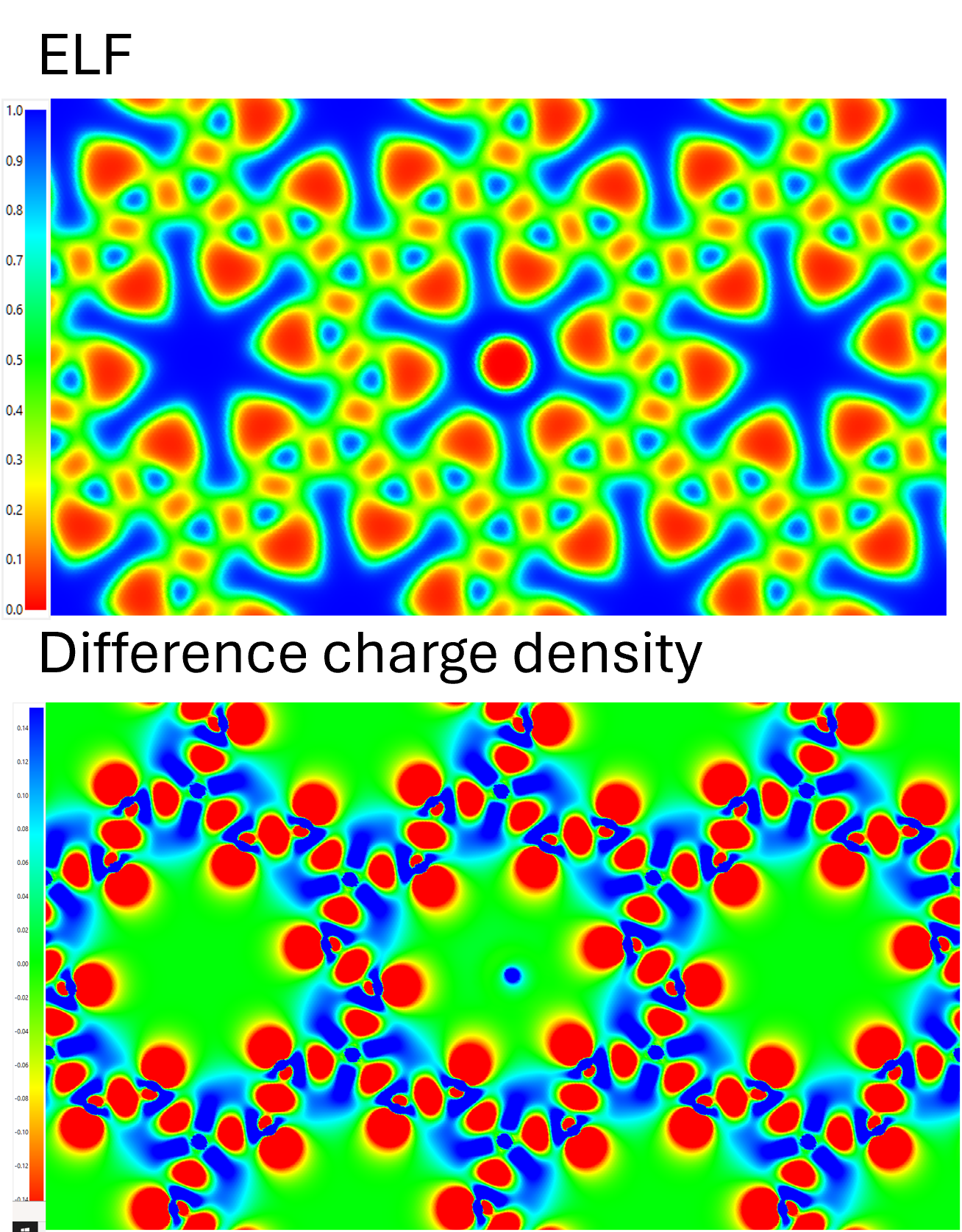} \\ (a)
\end{minipage}
\begin{minipage}{0.43\textwidth}
\includegraphics[width=\linewidth]{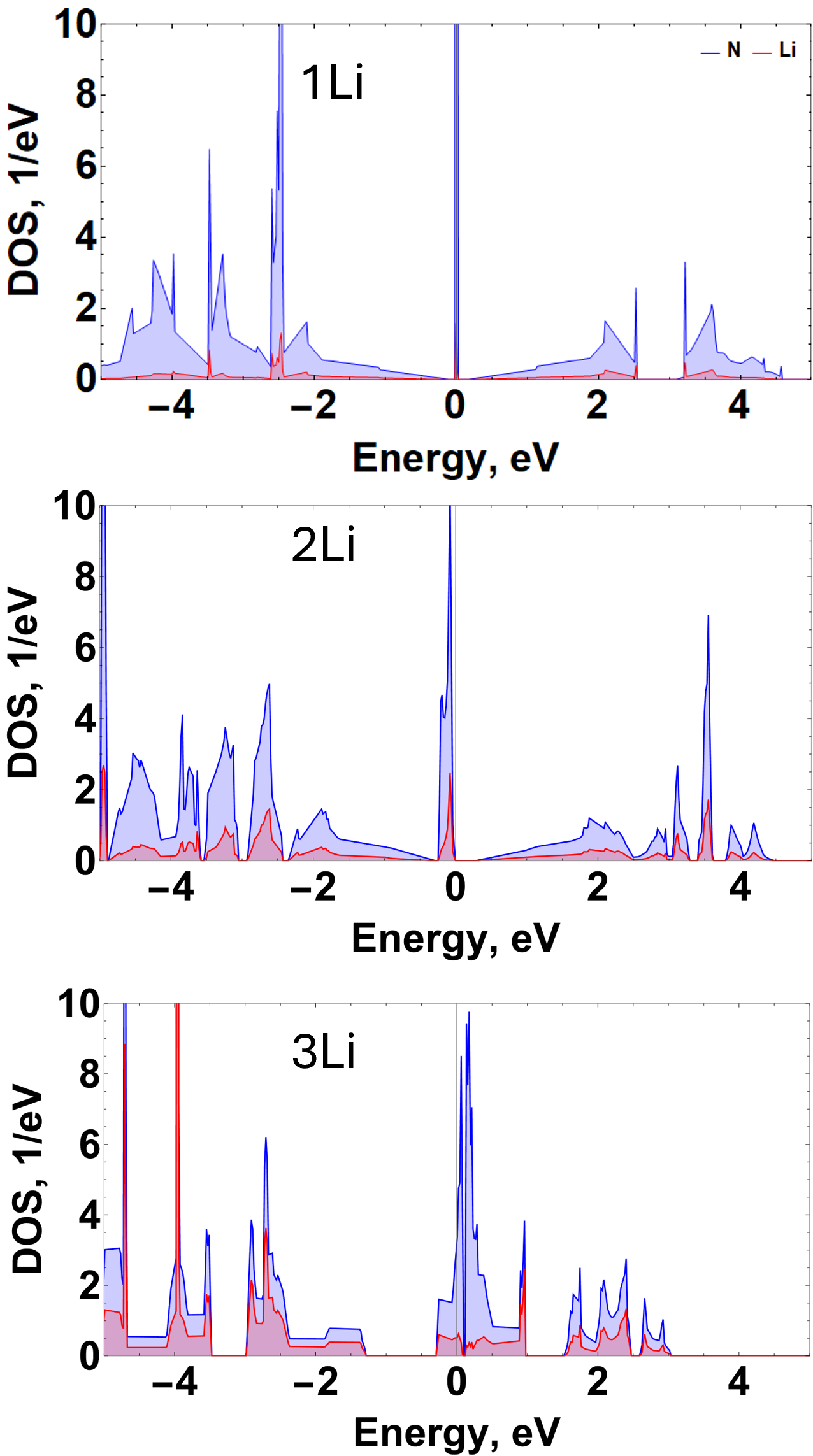} \\ (b)
\end{minipage}
\caption{ELF plot, difference charge density of 2D-N$_8$ with one Li-ion adsorbed on the 2$\times$2 cell and corresponding DOSes with one, two, and three adsorbed Li-ions.}
\label{fig_Li_electronic}
\end{figure}

The described exothermic adsorption process often results in the formation of metallic clusters. Indeed, during the adsorption of metal ions, the band gap completely closes as presented in Fig.\,\ref{fig_Li_electronic}. The DOS plot revealed the significant contribution at the Fermi level from both Li and N \textit{p}-electrons. In comparison with Fig.\,\ref{fig_electronic}\,b for bare N$_8$ density of states, it follows that the addition of the metal ion leads to an energy down-shift of the sharp peak at the bottom of the conduction band, leading to a metallic behavior.

\begin{figure}[h!]
\centering
\begin{minipage}{0.49\textwidth}
\includegraphics[width=1\linewidth]{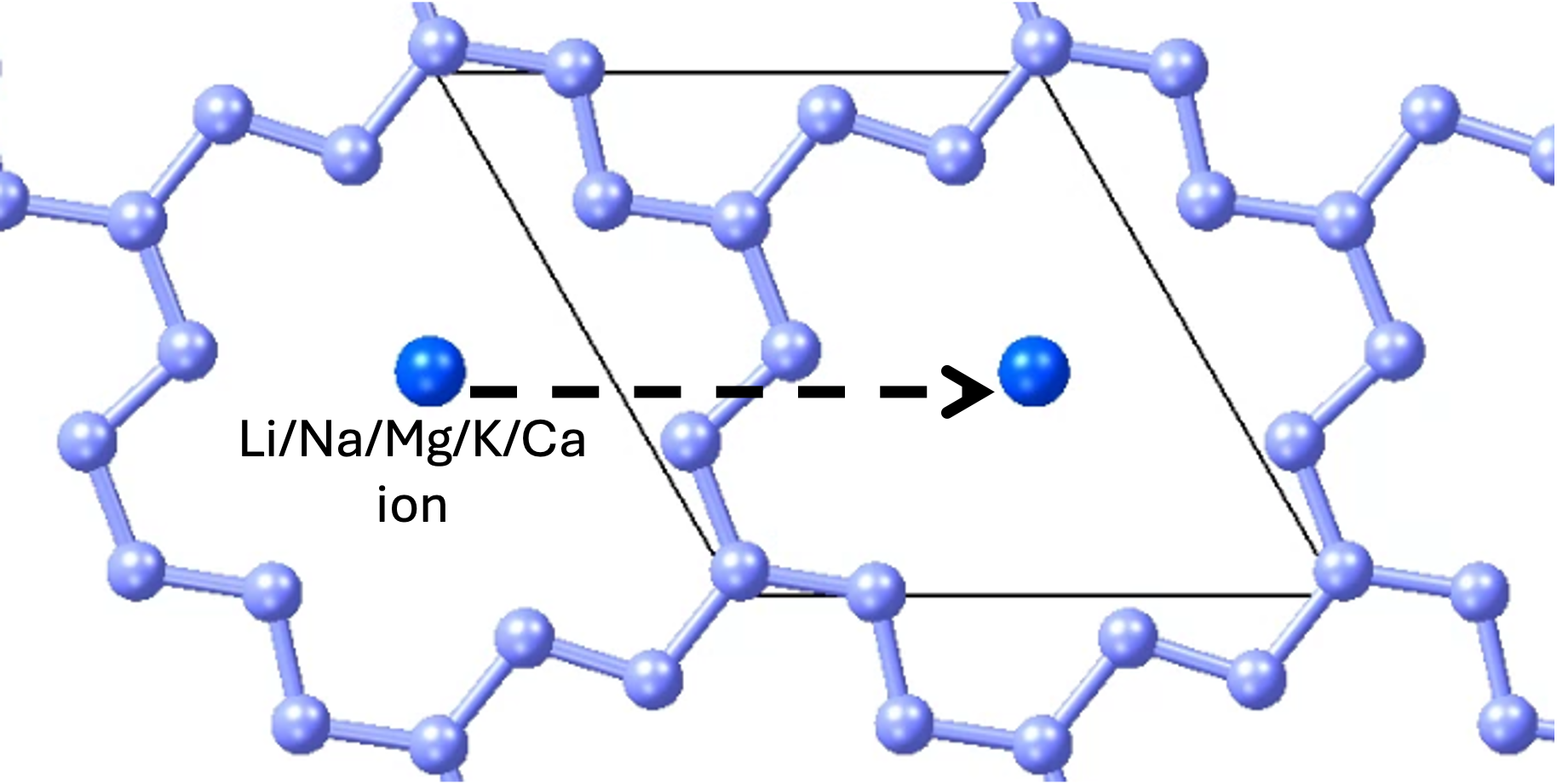}
\end{minipage}
\begin{minipage}{0.49\textwidth}
\includegraphics[width=1\linewidth]{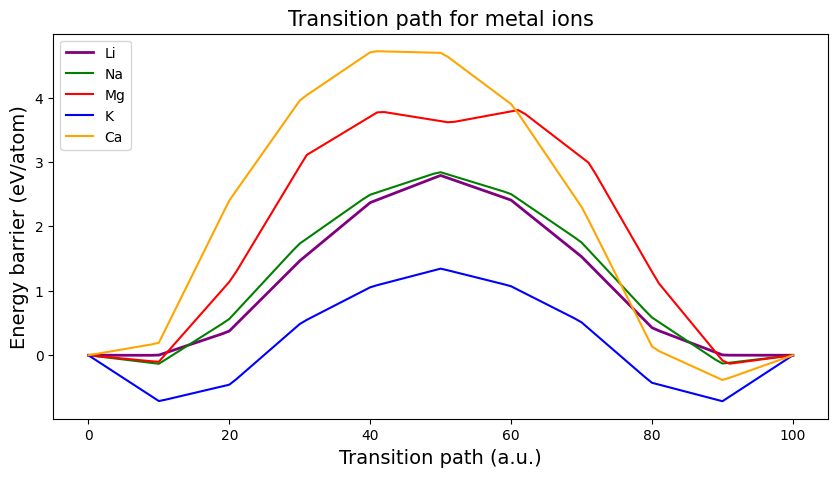}
\end{minipage}
\caption{Diffusion energy barriers for 2D-N$_8$ monolayer along the indicated path.}
\label{N8_diffusion}
\end{figure}

The calculation of the diffusion barrier is an essential characteristic in evaluating the performance and cycling efficiency of the charge/discharge process. Considering the most stable adsorption site for one metal ion per unit cell (shown in Fig.\,\ref{fig_cell}) the diffusion mechanism was investigated by calculating the migration path of metal ions from one pore to the nearest neighbor by crossing the N-N bond slightly above it and demonstrated in Fig.\,\ref{N8_diffusion}. 
The energy barriers obtained are 2.79, 2.84, 3.80, 1.3, and 4.72\,eV for Li, Na, Mg, K, and Ca ions, respectively. These values are higher than those of other 2D cathode materials, revealing potential limitations in achieving fast discharging rates. However, there is a range of works in which substitutions in the host matrix can significantly reduce energy barriers and improve other battery performance characteristics. That assumption will be examined in the next publications.

\subsection{Summary}
\label{summary}

This study thoroughly investigated the geometrical structure, stability, storage capacity, electronic properties, diffusivity, mechanical characteristics, and adsorption characteristics of the 2D N$_8$ monolayer using density functional theory (DFT) simulations. We aimed to evaluate its potential as an electrode material for next-generation post-lithium batteries.

To begin, we performed a comprehensive stability assessment. This meticulous analysis covered multiple aspects, revealing robust performance in several key areas:\\
\textbf{Mechanical Resilience:} The structure exhibited remarkable durability under various stress conditions, confirming its ability to maintain integrity throughout the cycle.\\
\textbf{Dynamic Stability:} We analyzed the vibrational modes and found no imaginary frequencies for layered N$_8$ structure, indicating stability during dynamic processes.\\
\textbf{Electronic Stability:} Electronic structure calculations showed consistent band gap values, suggesting stability in electronic properties under different environmental conditions.\\
\textbf{Thermodynamic Viability}: Formation energy calculations confirmed that the N$_8$ configuration is energetically favorable compared to competing structures, ensuring its viability as an electrode material.\\
\textbf{Thermal Resilience:} Thermal analysis revealed that the material can withstand elevated temperatures (up to 500\,K) without significant degradation, which makes it suitable for practical applications.

This comprehensive stability check indicates that the N$_8$ monolayer excels across all examined dimensions, showcasing its promise for high-performance energy storage systems.

Next, we evaluated the quantitative characteristics of the material. Our findings demonstrated that the two-dimensional N$_8$ structure can form stable adsorption complexes with various metal ions, achieving an impressive theoretical specific capacity (TSC) of 1675 mAh/g for sodium (Na) and magnesium (Mg) ions. This capacity corresponds to the adsorption of up to seven ions on one side of the monolayer. It highlights the material's exceptional storage capabilities and versatility in accommodating different metallic ions, making it a strong candidate for enhancing battery performance.

Furthermore, the averaged open circuit voltages (OCVs) for the examined metal ions ranged from 2.05 to 3.36 eV, indicating their suitability for cathode applications. Finally, our diffusion barrier calculations revealed a minimal barrier of 1.3 eV for potassium ions (K). This suggests that further substitutions could be explored to reduce barriers and enhance the N$_8$ monolayer's performance characteristics.

\section*{Acknowledgments}
The work of I. Piyanzina (Gumarova) was supported by the grant 24PostDoc/2-2F006.

\end{document}


\begin{frontmatter}



\title{A DFT study on 18-crown-6-like-N$_8$ structure as a  material for metal-ions storage: stability and performance}

\author{Irina I. Piyanzina $^{a,b*}$, Regina M. Burganova$^{b}$, Sadegh Kaviani$^{b}$, Oleg V. Nedopekin$^{b}$, Hayk Zakaryan$^{a}$  \\ 
irina.gumarova@ysu.am}

\affiliation{organization={Computational Materials Science Laboratory, Center of Semiconductor Devices and Nanotechnology, Yerevan State University, Republic of Armenia},    
addressline={1 Alex Manoogian St.}, 
           city={Yerevan},
            postcode={0025}, 
          country={Republic of Armenia}}

\affiliation{organization={Institute of Physics, Kazan Federal University},
            addressline={16~Kremlyovskaya~str.}, 
            city={Kazan},
            postcode={420008}, 
           country={Russia}}

\end{frontmatter}




\begin{figure}[h!] 
\centering
\begin{minipage}{\textwidth}
\includegraphics[width=\linewidth]{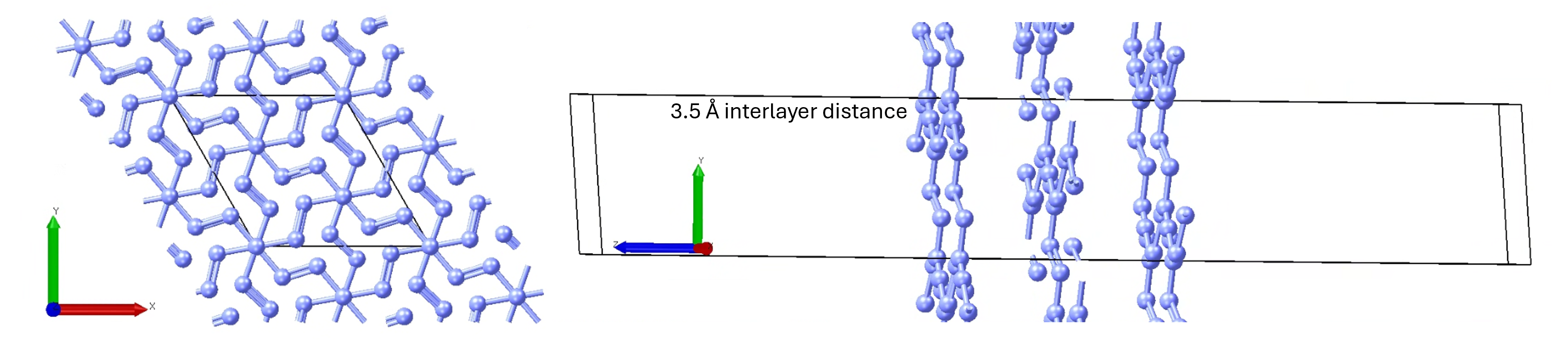} \\ (a)
\end{minipage}
\caption{Three N$_8$ monolayers surrounded by 23\,\AA~vacuum region.}
\label{fig_phonon}
\end{figure}

\begin{figure}[h!] 
\centering
\begin{minipage}{0.49\textwidth}
\includegraphics[width=\linewidth]{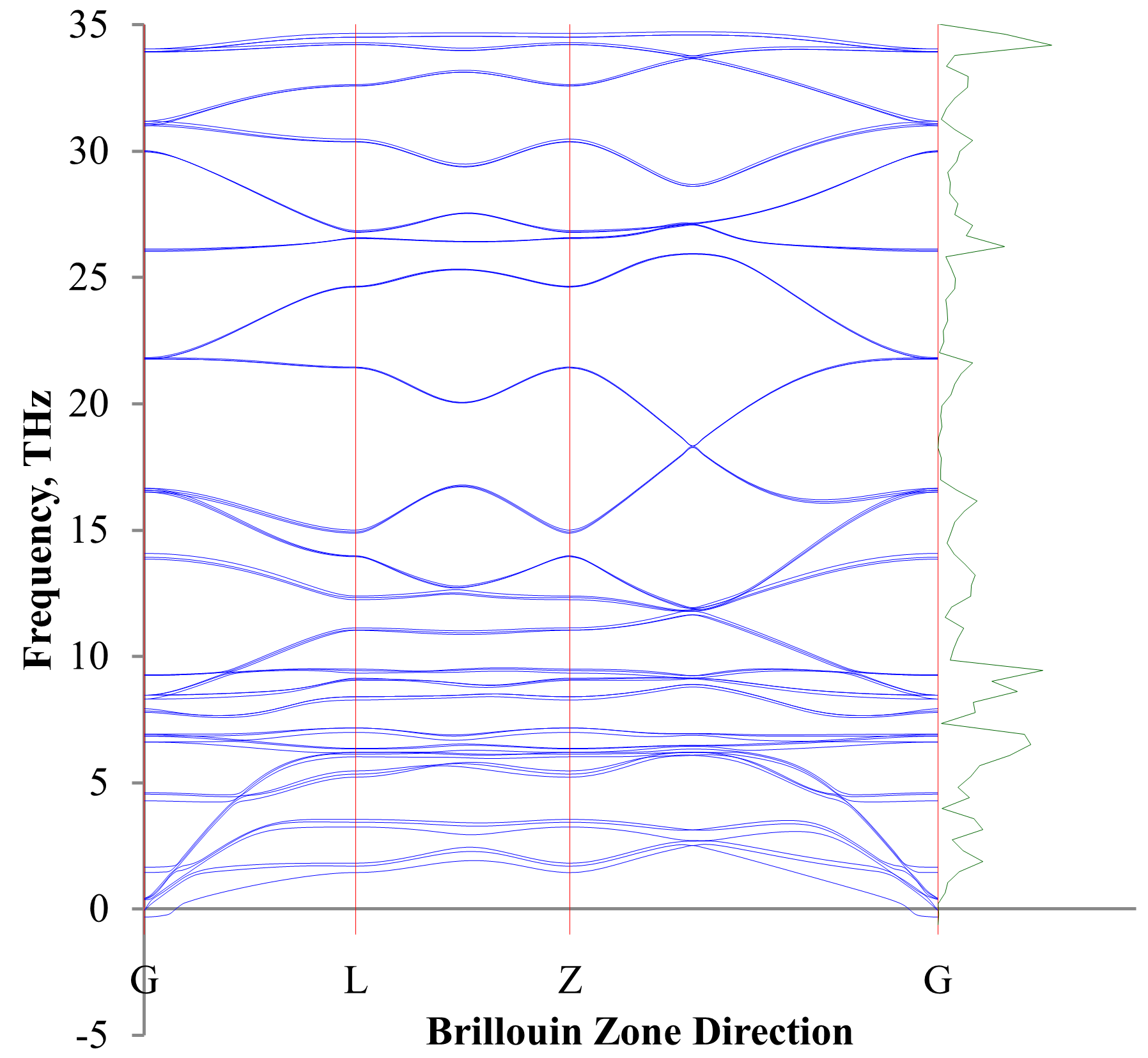} \\ (a)
\end{minipage}
\begin{minipage}{0.49\textwidth}
\includegraphics[width=\linewidth]{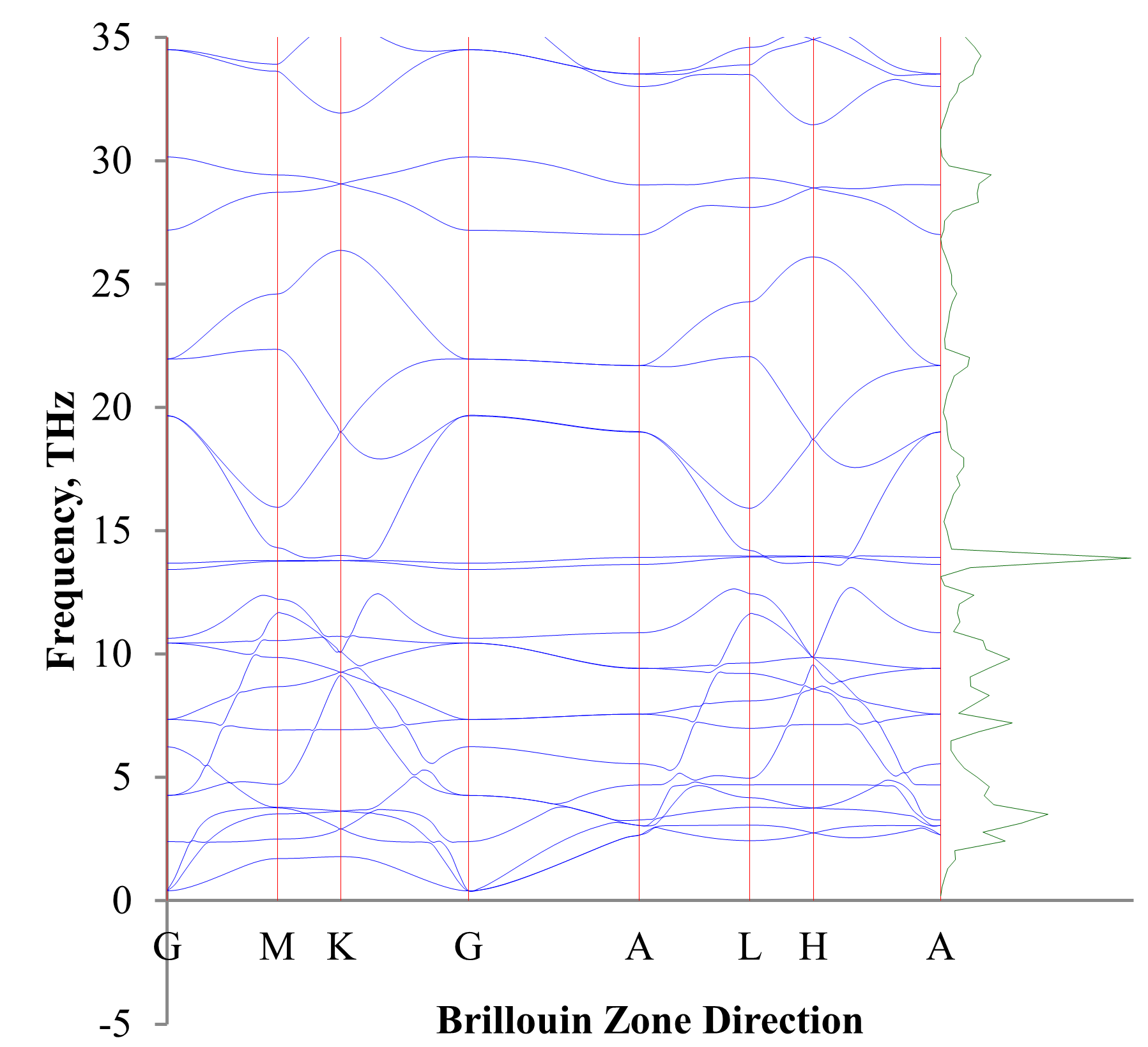} \\ (b)
\end{minipage}
\caption{(a) Phonon dispersion curves of the of pristine three N$_8$ monolayers depicted in Fig.\,1 of Supplementary and (b) of a single monolayer with an adsorbed metal ion.}
\label{fig_phonon}
\end{figure}

\begin{figure}[h!]
\centering
\begin{minipage}{0.32\textwidth}
\includegraphics[width=\linewidth]{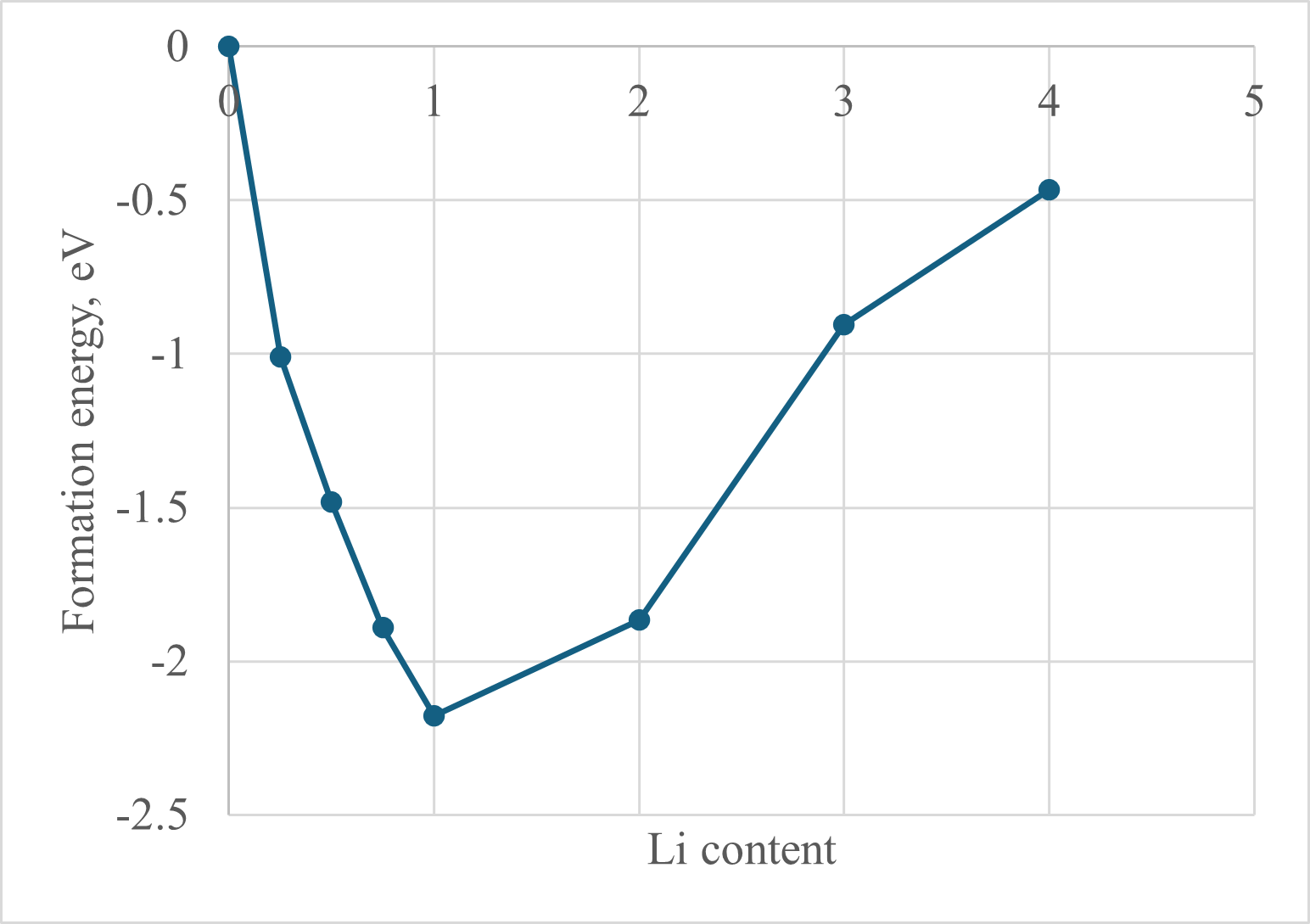} \\ Li
\end{minipage}
\begin{minipage}{0.32\textwidth}
\includegraphics[width=\linewidth]{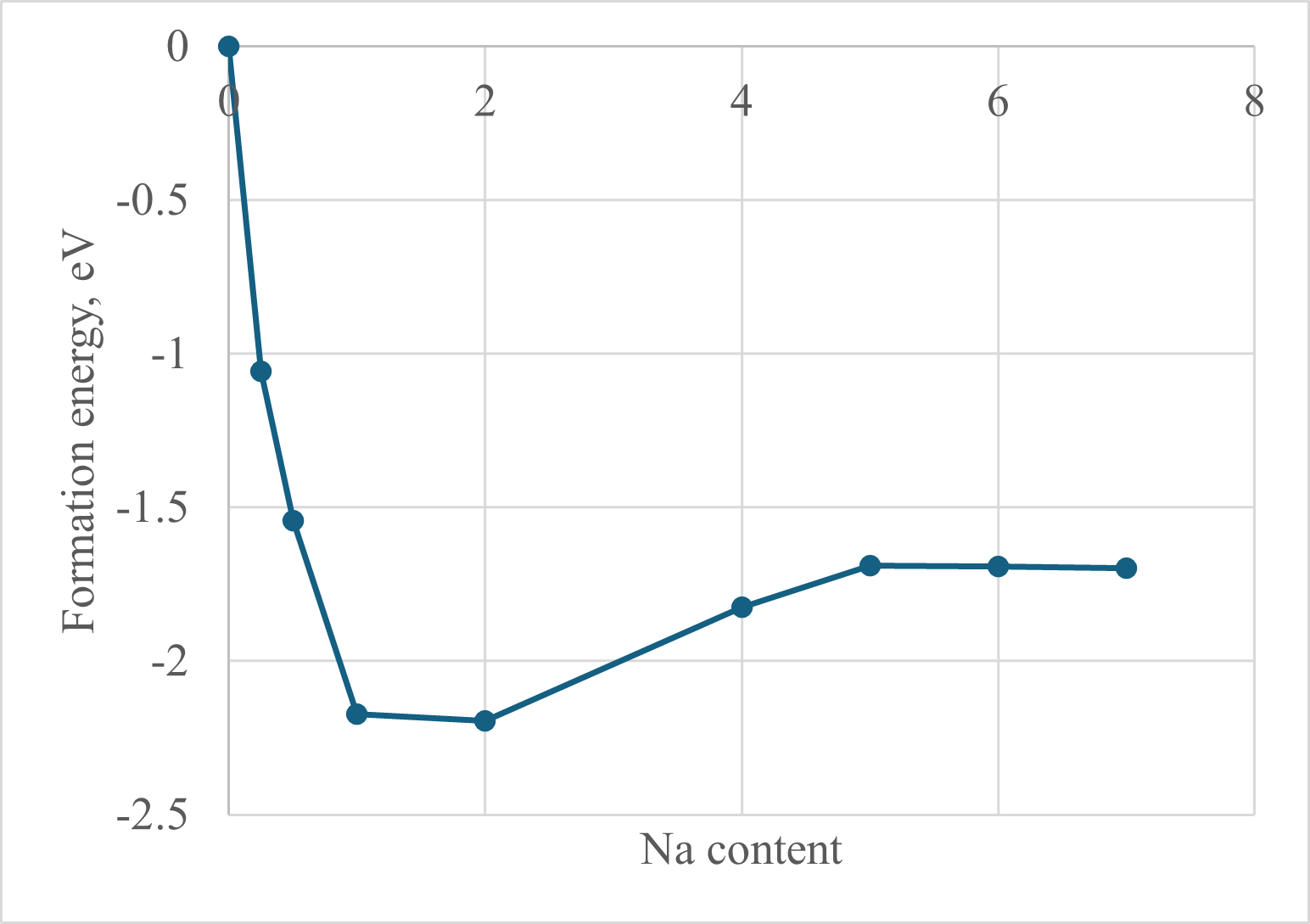} \\ Na
\end{minipage}
\begin{minipage}{0.32\textwidth}
\includegraphics[width=\linewidth]{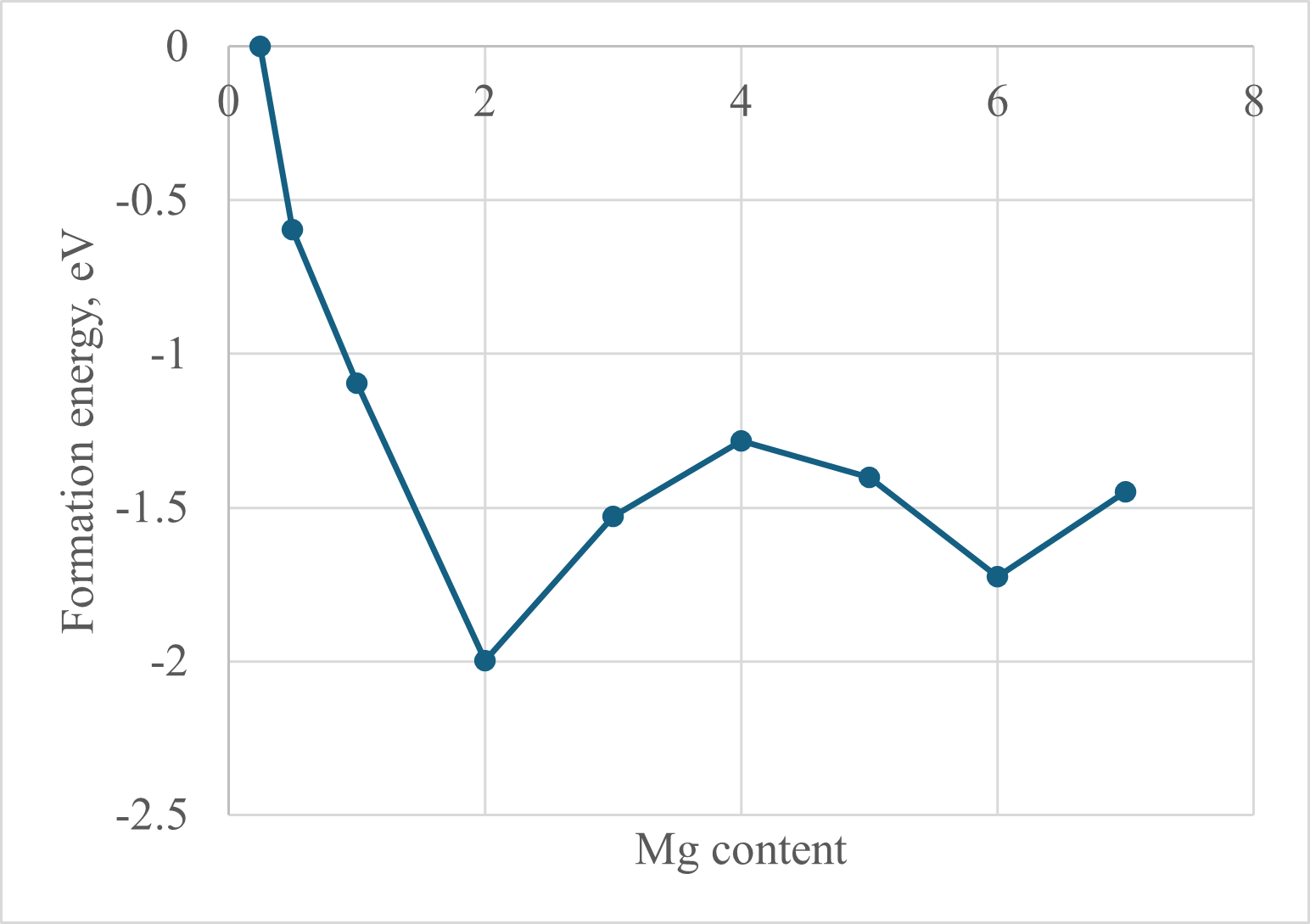}  \\ Mg
\end{minipage}
\begin{minipage}{0.32\textwidth}
\includegraphics[width=\linewidth]{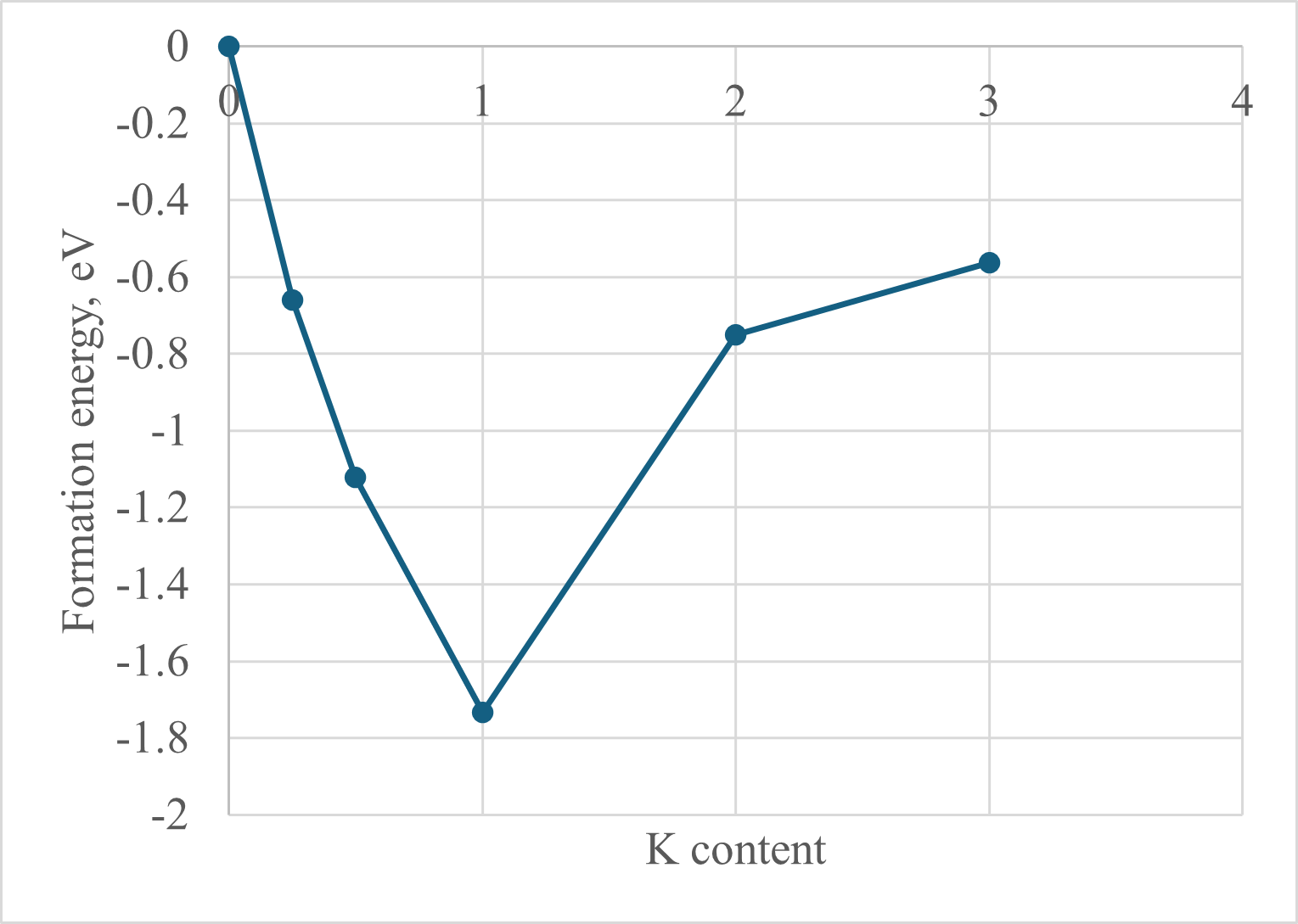}  \\ K
\end{minipage}
\begin{minipage}{0.32\textwidth}
\includegraphics[width=\linewidth]{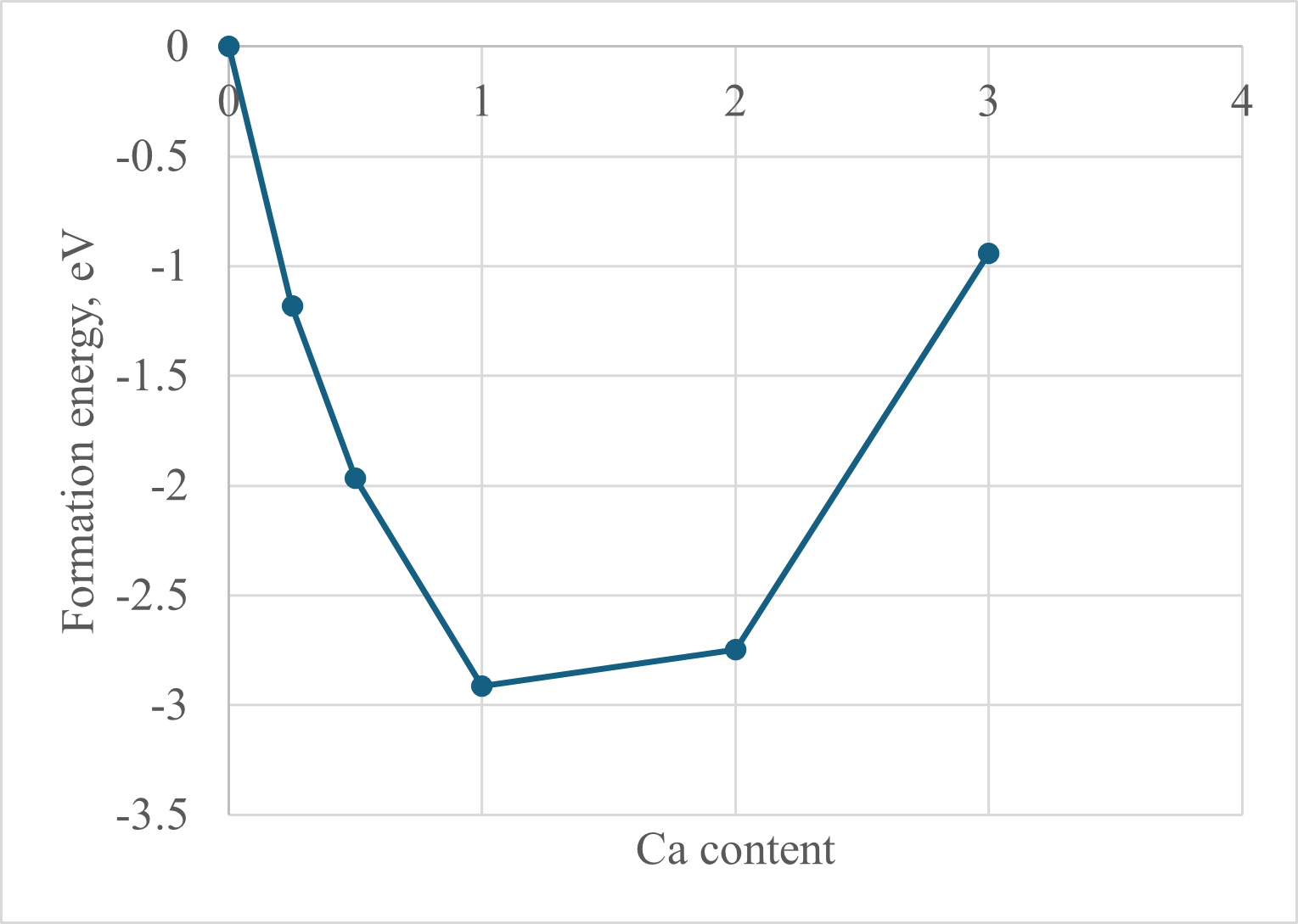}  \\ Ca
\end{minipage}
\caption{Formation energies for pristine 2D-N$_8$ monolayer with Li/Na/Mg/K/Ca adsorbed metal ions with different content.}
\label{fig_hull}
\end{figure}

\begin{figure}[h!]
\centering
\begin{minipage}{0.32\textwidth}
\includegraphics[width=\linewidth]{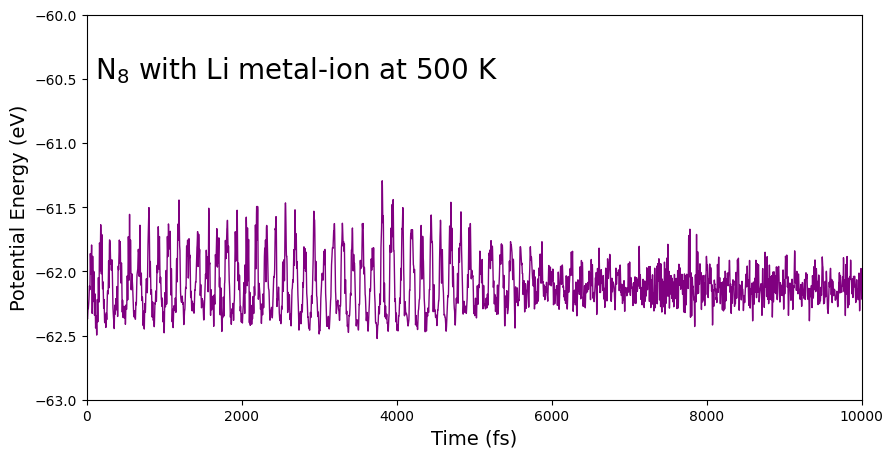} \\ Li
\end{minipage}
\begin{minipage}{0.32\textwidth}
\includegraphics[width=\linewidth]{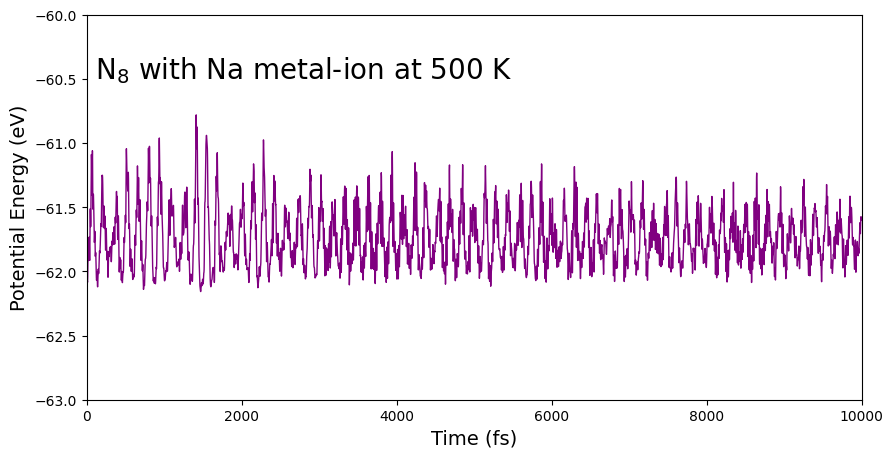} \\ Na
\end{minipage}
\begin{minipage}{0.32\textwidth}
\includegraphics[width=\linewidth]{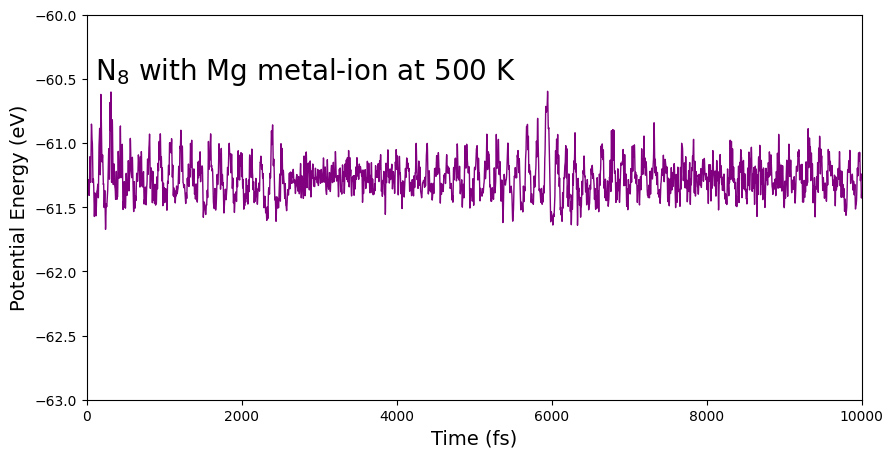}  \\ Mg
\end{minipage}
\begin{minipage}{0.32\textwidth}
\includegraphics[width=\linewidth]{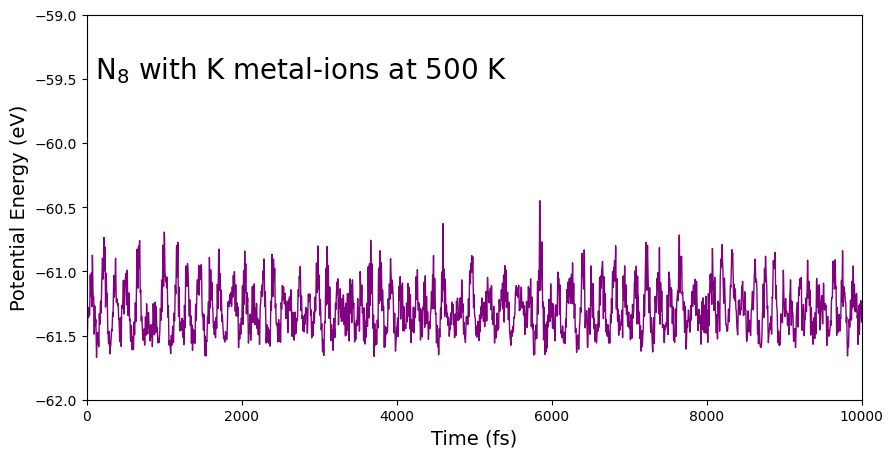}  \\ K
\end{minipage}
\begin{minipage}{0.32\textwidth}
\includegraphics[width=\linewidth]{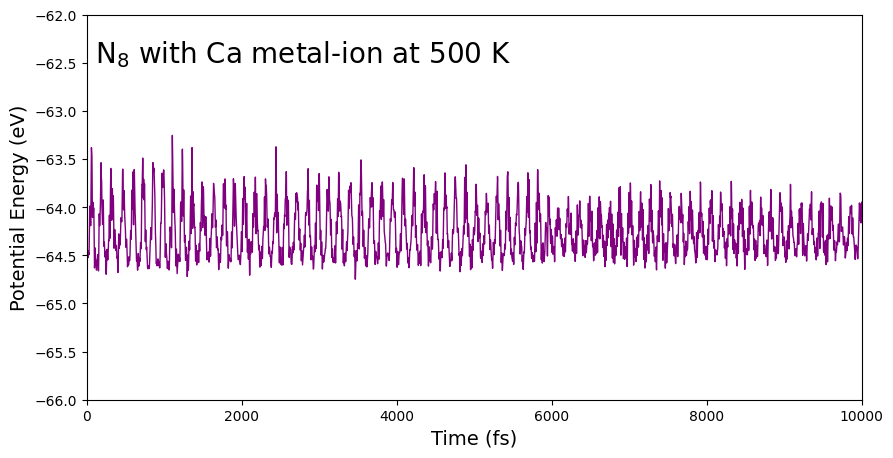}  \\ Ca
\end{minipage}
\caption{AIMD simulation results for Li/Na/Mg/K/Ca  on pristine N$_8$ monolayer.}
\label{fig_aimds}
\end{figure}